\documentclass[aps,prd,
comment, 
superscriptaddress,groupedaddress,
amsmath,amssymb,floatfix,footinbib]
{revtex4}  
\usepackage{float}
\usepackage{graphicx}
\usepackage{bm}
\usepackage[usenames,dvipsnames]{color}
\usepackage{slashed}
\usepackage{hyperref}
\usepackage{slashed}
\usepackage{graphics}
\usepackage{multirow}
\usepackage{tablefootnote}
\usepackage{booktabs}
\usepackage{amsmath}
\usepackage[capitalise]{cleveref}

\begin{document}

\renewcommand{\(}{\left(}
\renewcommand{\)}{\right)}
\renewcommand{\{}{\left\lbrace}
\renewcommand{\}}{\right\rbrace}
\renewcommand{\[}{\left\lbrack}
\renewcommand{\]}{\right\rbrack}
\renewcommand{\Re}[1]{\mathrm{Re}\!\{#1\}}
\renewcommand{\Im}[1]{\mathrm{Im}\!\{#1\}}
\newcommand{\dd}[1][{}]{\mathrm{d}^{#1}\!\!\;}
\newcommand{\del}{\partial}
\newcommand{\nn}{\nonumber}
\newcommand{\ie}{i.e.\,}
\newcommand{\cf}{cf.\,}
\newcommand{\refeq}[1]{Eq.~(\ref{eq:#1})}
\newcommand{\refeqs}[2]{Eqs.~(\ref{eq:#1})-(\ref{eq:#2})}
\newcommand{\reffig}[1]{Fig.~\ref{fig:#1}}
\newcommand{\refsec}[1]{Section \ref{sec:#1}}
\newcommand{\reftab}[1]{Table \ref{tab:#1}}
\newcommand{\order}[1]{\mathcal{O}\({#1}\)}
\newcommand{\fv}[1]{\left(\begin{array}{c}#1\end{array}\right)}%

\def\tcb#1{\textcolor{blue}{#1}}
\def\tcr#1{\textcolor{red}{#1}}
\def\tcg#1{\textcolor{green}{#1}}
\def\tcc#1{\textcolor{cyan}{#1}}
\def\tcv#1{\textcolor{violet}{#1}}
\def\tcm#1{\textcolor{magenta}{#1}}
\def\tcpn#1{\textcolor{pink}{#1}}
\def\tcpr#1{\textcolor{purple}{#1}}
\definecolor{schrift}{RGB}{120,0,0}

\def \azeL{{H_0^L}}
\def \azeR{{H_0^R}}
\def \apaL{{H_\parallel^L}}
\def \apaR{{H_\parallel^R}}
\def \apeL{{H_\perp^L}}
\def \apeR{{H_\perp^R}}

\newcommand{\alphas}{\alpha_\mathrm{s}}
\newcommand{\alphae}{\alpha_\mathrm{e}}
\newcommand{\gfermi}{G_\mathrm{F}}
\newcommand{\GeV}{\,\mathrm{GeV}}
\newcommand{\MeV}{\,\mathrm{MeV}}
\newcommand{\amp}[1]{\mathcal{A}\left({#1}\right)}
\newcommand{\wilson}[2][{}]{\mathcal{C}_{#2}^{\mathrm{#1}}}
\newcommand{\bra}[1]{\left\langle{#1}\right\vert}
\newcommand{\ket}[1]{\left\vert{#1}\right\rangle}

\def\be{\begin{equation}}
\def\ee{\end{equation}}
\def\bea{\begin{eqnarray}}
\def\eea{\end{eqnarray}}
\def\bm{\begin{matrix}}
\def\em{\end{matrix}}
\def\bpm{\begin{pmatrix}}
    \def\epm{\end{pmatrix}}

{\newcommand{\lsim}{\mbox{\raisebox{-.6ex}{~$\stackrel{<}{\sim}$~}}}
{\newcommand{\gsim}{\mbox{\raisebox{-.6ex}{~$\stackrel{>}{\sim}$~}}}
\def\mpl{M_{\rm {Pl}}}
\def\gev{{\rm \,Ge\kern-0.125em V}}
\def\tev{{\rm \,Te\kern-0.125em V}}
\def\mev{{\rm \,Me\kern-0.125em V}}
\def\ev{\,{\rm eV}}

\def\Mpl{M_{\rm Pl}}
\def\mpl{m_{\rm pl}}

\def\Lb{{\Lambda_b}}
\def\Ld{{\Lambda}}
\def\Lst{{\Lambda^\ast}}
\def\mLb{{m_{\Lambda_b}}}
\def\mmLb{{m^2_{\Lambda_b}}}
\def\mL{{m_{\Lambda}}}
\def\mmL{{m^2_{\Lambda}}}

\def\mLst{{m_{\Lambda^\ast}}}
\def\mmLst{{m^2_{\Lambda^\ast}}}

\title{\boldmath  \color{schrift}{$\Lambda_b \to \Lambda^{(\ast)}\nu\bar{\nu}$ decays and the recent Belle-II $B^+\to K^+\nu\bar{\nu}$ data}}
\author{Diganta Das}
\affiliation{Center for Computational Natural Sciences and Bioinformatics,
International Institute of Information Technology, Hyderabad 500 032, India}

\author{Dargi Shameer}
\affiliation{Center for Computational Natural Sciences and Bioinformatics,
International Institute of Information Technology, Hyderabad 500 032, India}

\author{Ria Sain}
\affiliation{Institute of Particle Physics and Key Laboratory of Quark and Lepton Physics (MOE), Central
China Normal University, Wuhan, Hubei 430079, China}

\begin{abstract}
The Belle-II experiment has recently reported the first measurement of $B^+ \to K^+ \nu\bar{\nu}$ decay which exceeds the Standard Model prediction by approximately 2.7$\sigma$. The deviation may indicate the presence of new physics beyond the Standard Model in the $b\to s\nu\bar{\nu}$ sector. Under this assumption, we study the hadronic $\Lambda_b \to \Lambda(\to p\pi)\nu\bar{\nu}$ and $\Lambda_b \to \Lambda^\ast(\to N\!\bar{K})\nu\bar{\nu}$ decays within both the Standard Model and beyond. We work in a low energy effective field theory framework with additional light right-handed neutrinos. We calculate the differential branching ratios of these decay modes and explore the implications of the Belle-II results through various observables.

\end{abstract}
\maketitle

\section{Introduction}
The flavor-changing neutral current (FCNC) transitions such as the $b\to s\ell^+\ell^-$ and the $b\to s\nu\bar{\nu}$ are among the most sensitive probe for physics beyond the Standard Model (SM). Semileptonic $B$-decays mediated by the $b\to s\ell^+\ell^-$ transition have been extensively studied in experiments which have ultimately led to a good agreement with the SM predictions, further constraining the landscape of physics beyond the Standard Model (BSM) physics in this sector. A key distinction between the two transitions is that unlike the $b\to s\ell^+\ell^-$, the $b\to s\nu\bar{\nu}$ is free from sizable electromagnetic contribution from the charm quarks, and the photon pole contribution. Therefore, the $b\to s\nu\bar\nu$ transitions are theoretically better understood, but remain experimentally challenging to detect due to the presence of neutrinos in the final state. Recently, the Belle-II collaboration has presented the first evidence of the rare $B^+\to K^+\bar{\nu} \nu$ decay using two different measurement methods: a standard hadronic tag (had), and a novel inclusive tag (incl) \cite{Belle-II:2023esi}. The branching ratios that the methods yield are
\begin{eqnarray}
\begin{split}
    {\rm Br}(B^+ \to K^+\bar{\nu} \nu)_{\rm had} &=& (1.1^{+0.9+0.8}_{-0.8-0.5}) \times 10^{-5}\, ,\\
    {\rm Br}(B^+ \to K^+\bar{\nu}\nu)_{\rm incl} &=& (2.7 \pm 0.5 \pm 0.5) \times 10^{-5}\, ,
\end{split}
\end{eqnarray}
where the first and the second uncertainties are statistical and systematic, respectively. The combination of the two measurements
\begin{equation}\label{BelleII-23}
    {\rm Br}(B^+ \to K^+ \bar{\nu}\nu)_{\rm exp} = (2.3 \pm 0.7) \times 10^{-5}\, ,
\end{equation}
is in excess of 2.7$\sigma$ compared to the SM predictions \cite{Altmannshofer:2009ma,Buras:2014fpa}
\begin{equation}
    {\rm Br} (B^+ \to K^+\bar{\nu}\nu)_{\rm SM} = (4.29 \pm 0.23) \times 10^{-6}\, .
\end{equation}
The result \eqref{BelleII-23} is also larger than the previous Belle-II combination ${\rm Br}(B^+\to K^+\nu\bar{\nu})_{\rm exp21} = (1.1 \pm 0.4) \times 10^{-5}$ by almost a factor of two but the two results remain consistent within uncertainties. 

The $b\to s\nu\bar{\nu}$ transitions also mediate the $B^0\to K^{\ast 0}\nu\bar{\nu}$ decay and the upper limit on the branching ratio from the Belle collaboration is ${\rm Br} (B^0\to K^{\ast 0}\nu\bar{\nu})_{\rm exp} < 1.8 \times 10^{-5}$ \cite{Belle:2017oht}. The excess over the SM prediction can be expressed in terms of ratios \cite{He:2023bnk}
\begin{eqnarray}\label{eq:Belle-II-ratios}
\begin{split}
    R^K_{\nu\nu} &=& \frac{{\rm Br}(B^+\to K^+\nu\bar{\nu})_{\rm exp}}{{\rm Br}(B^+\to K^+\nu\bar{\nu})_{\rm SM}} = 5.4 \pm 1.6\, ,\\
    R^{K^\ast}_{\nu\nu} &=& \frac{{\rm Br}(B\to K^{\ast}\nu\bar{\nu})_{\rm exp}}{{\rm Br}(B\to K^{\ast}\nu\bar{\nu})_{\rm SM}} < 2.7\quad \text{or}  \quad 1.9\, ,
    \end{split}
\end{eqnarray}

While BSM interpretations can be invoked, further scrutiny, both experimental and theoretical, is required to either confirm or resolve the discrepancies. In this context the semi-leptonic decays of $\Lambda_b$ baryons will play important role. Experimental state-of-the-art for $\Lambda_b$ decays are not as matured as its mesonic counterpart, but progress is being made. Following the observation of $\Lambda_b\to \Lambda\mu^+\mu^-$ by the CDF Collaboration \cite{CDF:2011buy}, the LHCb Collaboration observed the decay in \cite{LHCb:2013uqx} and subsequently presented their angular analysis in \cite{LHCb:2015tgy, LHCb:2018jna}. The $\Lambda_b \to \Lambda^{(\ast)}\nu\bar{\nu}$ modes (including $B_s^0\to \phi\nu\bar{\nu}$ modes) can only be measured in a Tera-Z factory such as FCC-ee \cite{FCC:2018evy, Bernardi:2022hny}. This will provide a new avenue to explore physics within and beyond the SM in $b \to s\nu\nu$. 

Previously, the $\Lambda_b\to \Lambda \nu\bar{\nu}$ mode has been studied in the SM with polarized $\Lambda_b$ baryon \cite{Chen:2000mr, Altmannshofer:2025eor}, in the context of unparticle physics \cite{Aliev:2007rm}, and in a leptophobic $Z^\prime$ model \cite{Sirvanli:2007yq, Das:2023kch}. In this paper we are interested in the implications of the recent Belle-II results on the  $\Lambda_b \to \Lambda(\to p\pi)\nu\bar{\nu}$ and $\Lambda_b \to \Lambda^\ast(\to N\!\bar{K})\nu\bar{\nu}$. Assuming physics beyond the SM to is responsible for the Belle-II data, we present bounds on the $\Lambda_b \to \Lambda(\to p\pi)\nu\bar{\nu}$ and $\Lambda_b \to \Lambda^\ast(\to N\!\bar{K})\nu\bar{\nu}$ observables under different new physics (NP) scenarios. The three scenarios considered are NP with lepton flavor universality (LFU), NP with LFU violation (LFUV) but conserved lepton flavor, and NP with lepton flavor violation (LFV). 

The paper is organized as follows: the effective Hamiltonian is presented in section \ref{sec:Effective-Hamiltonian}, which is followed by the discussions of observables in \cref{sec:observables}. The numerical implication of Belle-II data is presented in \cref{sec:belle-ii}. A summary is presented in \cref{sec:summary}.

\section{Effective Hamiltonian \label{sec:Effective-Hamiltonian} } 
Possible new physics at or above the electroweak scale entering the $b\to s\nu\bar{\nu}$ transition can be parameterised by a low energy effective theory  \cite{Grzadkowski:2010es, Jenkins:2017jig} at the $b$ mass scale involving dimension six operators. Keeping the possibility of light right-handed neutrinos, we work with an effective Hamiltonian
\begin{eqnarray}\label{eq:eff_Ham}
\mathcal{H}_{\rm eff} = -\frac{4G_{F}}{\sqrt{2}}\frac{\alpha_e}{4\pi} \lambda_t\bigg[C^{\rm SM} \mathcal{O}^{\rm SM}+ \sum_{A,B=L}^R \sum_{ij} C_{AB}^{ij} \mathcal{O}_{AB}^{ij} \bigg]\, ,
\end{eqnarray}
where $G_F$ is the Fermi-constant, $\alpha_e=e^2/4\pi$ is the fine structure constant, and $\lambda_t=V_{tb}V^\ast_{ts}$ are the Cabibbo-Kobayashi-Maskawa (CKM) elements. The SM operator reads
\be
O^{\rm SM} = (\bar{s}\gamma_\mu P_L b) (\bar{\nu} \gamma^\mu P_L \nu)\, .
\ee
The corresponding Wilson coefficient written in terms of a function of $x_t = m_t^2/m_W^2$ as \cite{Buchalla:1998ba, Brod:2010hi} 
\be\label{eq:SMWC}
C^{\rm SM} = -\frac{2X(x_t)}{ \sin^2\theta_w }\, ,\quad X(x_t) = 1.469 \pm 0.017\, .
\ee
Here $\theta_w$ is the Weinberg angle and the function $X(x_t)$ includes the NLO QCD corrections, and two-loop electroweak contributions. The lepton flavor specific NP operators are 
\begin{align}
&\mathcal{O}_{AB}^{ij} = (\bar{s}\gamma_\mu P_A b) (\bar{\nu}^i \gamma^\mu P_B \nu^j)\, ,
\end{align}
where $P_{L,R} = (1\mp \gamma_5)/2$ is the left chiral projectors. In the SM all the $C_{AB}^{ij}$ vanish but they can be non-zero in physics beyond the SM.

\section{Observables \label{sec:observables}}
\subsection{Lepton Flavor Conserving New Physics}
When the new physics is lepton flavor conserving the total (SM+NP) differential branching ratios for the $\Lambda_b \to \Lambda(\to p\pi)\nu\bar{\nu}$ and the $\Lambda_b \to \Lambda^\ast(\to N\!\bar{K})\nu\bar{\nu}$ decays in terms of dilepton invariant mass squared $q^2$ and cosine of hadronic angle $\theta_{\Lambda^{(\ast)}}$ are \cite{Das:2018sms, Das:2018iap, Das:2020cpv} 
\bea\label{eq:Lambda-dqdcos}
\frac{d^2\mathcal{B}_\Lambda}{dq^2 d\cos\theta_\Lambda} &=&  N_{\Lambda}^2 \bigg[ \bigg(\bigl(m_{\Lambda_b}-m_{\Lambda}\bigr)^2s_{+}f_0^{A^2}+2q^2s_{+}f_{\perp}^{A^2} \bigg) \mathbb{C}_- + \bigg(\bigl(m_{\Lambda_b}+m_{\Lambda}\bigr)^2s_{-}f_0^{V^2}+2q^2s_{-}f_{\perp}^{V^2} \bigg) \mathbb{C}_+ \,\nn\\ & - & 2 \cos{\theta_\Lambda} \alpha_\Lambda\bigg( 2 q^2 \sqrt{s_- s_+}f^A_{\perp}f^V_{\perp} +   \bigl(\mmLb -\mmL \bigr) \sqrt{s_- s_+}f^A_{0}f^V_{0} \bigg)  \mathsf{Re}(\mathbb{C}^\prime) \bigg]\, ,\\
\label{eq:LamSt-dq}
\frac{d^2\mathcal{B}_{\Lambda^\ast}}{dq^2 d\cos\theta_\Lst} &=& \frac{N_{\Lambda^\ast}^2}{12m_{\Lambda^\ast}^2} \bigg[ \bigg( (1+3\cos^2\theta_{\Lst}) \big(f_0^{A^2}s_+^2s_-\bigl(m_{\Lambda_b}-m_{\Lambda^\ast}\bigr)^2 + 2f_{\perp}^{A^2}q^2s_-s_+^2\big) +18(1-\cos^2\theta_{\Lst})f_g^{A^2}m^2_\Lst q^2s_-\biggr) \mathbb{C}_- \, \nn\\ & + &  \biggl((1+3\cos^2\theta_{\Lst}) \big(f_0^{V^2}s_-^2s_+\bigl(m_{\Lambda_b}+m_{\Lambda^\ast}\bigr)^2 + 2f_{\perp}^{V^2}q^2s_+s_-^2\big) +18(1-\cos^2\theta_{\Lst})f_g^{V^2}m^2_\Lst q^2s_+\biggr) \mathbb{C}_+ \bigg]\, , \quad\quad
\eea
where $\alpha_\Lambda=0.642\pm 0.013$ \cite{ParticleDataGroup:2016lqr} is a parameter related to the weak decay of $\Lambda$ to $p\pi^-$ \cite{Das:2018iap}, the normalization constants for $X=\Lambda,\Lambda^\ast$ are
\be
N_X^2 = \tau_{\Lambda_b} \frac{G_F^2 \lambda_t^2 \alpha_e^2 \sqrt{\lambda(\mmLb,m^2_X,q^2)}}{3.2^{11} m_{\Lambda_b}^3\pi^5}\, ,
\ee
where $ \tau_{\Lambda_b}$ is the $\Lambda_b$ lifetime \cite{ParticleDataGroup:2024cfk}, and
\be
s_\pm = (m_\Lb \pm m_X)^2 - q^2\, .
\ee
The interference between the new physics and the SM is aptly captured by the Wilson coefficients 
\bea\label{eq:nonLFV}
\mathbb{C}_- &=&\sum_{i} \bigg| C^{\rm SM}+ \big( C^{ii}_{LL}-C^{ii}_{RL}\big) \bigg|^2 + \sum_{i} \bigg|\big( C^{ii}_{LR}- C^{ii}_{RR} \big) \bigg|^2 \, ,\nn\\
\mathbb{C}_+ &=&\sum_{i} \bigg| C^{\rm SM}+ \big(C^{ii}_{LL}+C^{ii}_{RL}\big) \bigg|^2 + \sum_{i} \bigg| \big(C^{ii}_{LR}+C^{ii}_{RR}\big) \bigg|^2\, ,\\\quad\quad
\mathbb{C}^\prime &=& \sum_{i} \bigl(C^{\rm SM}+ \big(C^{ii}_{LL}-C^{ii}_{RL}\big)\bigr)^\ast  \times \bigl(C^{\rm SM}+\big(C^{ii}_{LL}+C^{ii}_{RL}\big)\bigr) + \sum_{i} \bigl(\big(C^{ii}_{LR}-C^{ii}_{RR}\big)\bigr)^\ast \bigl(\big(C^{ii}_{LR}+C^{ii}_{RR}\big)\bigr)\, ,\nn
\eea
where a sum is implicit for repeated indices. The functions $f_0^{V,A}, f_\perp^{V,A}, f_g^{V,A}$ are the $q^2$-dependent form factors required to describe the $\Lambda_b\to \Lambda^{(\ast)}$ hadronic matrix elements (see Appendix \ref{HMA}). The parametrizations of the hadronic matrix elements are according to the helicty basis \cite{Feldmann:2011xf, Descotes-Genon:2019dbw}.

By integrating the two-fold differential branching ratios \eqref{eq:Lambda-dqdcos} and \eqref{eq:LamSt-dq} a couple of observables can be constructed. They are differential branching ratios $d\mathcal{B}_{\Lambda^{(\ast)}}/dq^2$, longitudinal polarization fractions $F_L^{\Lambda^{(\ast)}}$, and hadronic-side forward-backward asymmetries $A^{h,\Lambda^{(\ast)}}_{{\rm FB}}$. The later two observables are obtained from the two-fold differential rates in the following way \cite{Das:2017ebx}
\bea\label{eq:FL-def}
F_L^X &=& \frac{1}{d\mathcal{B}_X/dq^2} \int_{-1}^{+1} d\cos\theta_X \frac{d^2\mathcal{B}_X}{dq^2d\cos\theta_X} \frac{1}{2}(5\cos^2\theta_X-1) \,,\\
\label{eq:AFBh-def}
A^{h,X}_{{\rm FB}} &=& \frac{\bigg[\int_{-1}^0 - \int_0^1 \bigg] d\cos\theta_X \frac{d^2\mathcal{B}_X}{dq^2d\cos\theta_X}}{\int_{-1}^{+1} d\cos\theta_X \frac{d^2\mathcal{B}_X}{dq^2d\cos\theta_X}} \, .
\eea
For the $\Lambda_b\to\Lambda(\to p\pi)\nu\bar{\nu}$ mode, the longitudinal polarization fraction reads 
\be
F_L^{\Lambda} = \frac{1}{3}\, ,
\ee
and the hadronic side forward-backward asymmetry is 
\bea
A^{h,\Lambda}_{\rm FB} = \frac{\alpha_\Lambda\bigg( 2 q^2 \sqrt{s_- s_+}f^A_{\perp}f^V_{\perp} +   \bigl(\mmLb -\mmL \bigr) \sqrt{s_- s_+}f^A_{0}f^V_{0} \bigg) \mathsf{Re}(\mathbb{C}^\prime) }{ \bigg(\bigl(m_{\Lambda_b}-m_{\Lambda}\bigr)^2s_{+}f_0^{A^2}+2q^2s_{+}f_{\perp}^{A^2} \bigg) \mathbb{C}_- + \bigg(\bigl(m_{\Lambda_b}+m_{\Lambda}\bigr)^2s_{-}f_0^{V^2}+2q^2s_{-}f_{\perp}^{V^2} \bigg) \mathbb{C}_+ }\, .
\eea
For the $\Lambda_b \to \Lambda^\ast(\to N\bar{K})\nu\bar{\nu}$ mode the longitudinal polarization fraction is
\be 
F_{L}^{\Lst} = \frac{2}{3} \frac{ \mathbb{C}_+ \mathbb{G}_1+\mathbb{C}_-\mathbb{G}_2 }{ \mathbb{C}_+\bigl(\mathbb{G}_1^{\prime}+ \mathbb{G}_1 \bigr)+\mathbb{C}_-\bigl(\mathbb{G}_2^{\prime}+ \mathbb{G}_2 \bigr)}\, ,
\ee
where $\mathbb{G}_1$ and $\mathbb{G}_2$ in terms of the $\Lambda_b\to\Lambda^\ast$ form factors read as
\bea
\mathbb{G}_1&=&(m_{\Lambda_b}+m_\Lst)^2f_0^{V^2}s_-^2s_++2f_\perp^{V^2}q^2s_-^2s_+\, , \nn \\
\mathbb{G}_2&=&(m_{\Lambda_b}-m_\Lst)^2f_0^{A^2}s_+^2s_-+2f_\perp^{A^2}q^2s_+^2s_-\, . \nn \\
\mathbb{G}_1^{\prime}&=&6f_g^{V^2}m^2_\Lst q^2s_+\, . \nn \\
\mathbb{G}_2^{\prime}&=&6f_g^{A^2}m^2_\Lst q^2s_-\, . \nn
\eea
The hadronic side forward-backward asymmetry vanishes since the $\Lambda^\ast$ decays through strong interaction. 

\subsection{Lepton Flavor Violating New Physics}
When the NP is lepton flavor violating (LFV), there is no interference between the SM and the NP. The new physics part of the differential branching ratios for the two modes are 
\bea\label{eq:LFV-dBdq2-Lam}
\frac{d^2\mathcal{B}_\Lambda}{dq^2 d\cos\theta_\Lambda} &=&  N_{\Lambda}^2 \bigg[ \bigg(\bigl(m_{\Lambda_b}-m_{\Lambda}\bigr)^2s_{+}f_0^{A^2}+2q^2s_{+}f_{\perp}^{A^2} \bigg) \mathbb{C}_-^{\rm LFV} + \bigg(\bigl(m_{\Lambda_b}+m_{\Lambda}\bigr)^2s_{-}f_0^{V^2}+2q^2s_{-}f_{\perp}^{V^2} \bigg) \mathbb{C}_+^{\rm LFV} \,\nn\\ & - & 2 \cos{\theta_\Lambda} \alpha_\Lambda\bigg( 2 q^2 \sqrt{s_- s_+}f^A_{\perp}f^V_{\perp} +   \bigl(\mmLb -\mmL \bigr) \sqrt{s_- s_+}f^A_{0}f^V_{0} \bigg)  \mathsf{Re}(\mathbb{C}^{\prime \rm LFV}) \bigg]\, ,\\
\label{eq:LFV-dBdq2-LamSt}
\frac{d^2\mathcal{B}_{\Lst}}{dq^2 d\cos\theta_\Lst }&=& \frac{N^2_\Lst}{12m_{\Lambda^\ast}^2}\biggl[\mathbb{C}_+^{\rm LFV}\biggl(18(1-\cos^2\theta_{\Lst})f_g^{V^2}m_\Lst^2q^2s_++(1+3\cos^2\theta_{\Lst})\bigl(f_0^{V^2}(m_{\Lambda_b}+m_{\Lst})^2 + 2f_\perp^{V^2}q^2\bigr)s^2_-s_+\biggr)\,\nn\\ & + & \mathbb{C}_-^{\rm LFV}\biggl(18(1-\cos^2\theta_{\Lst})f_g^{A^2}m_\Lst^2q^2s_-+(1+3\cos^2\theta_{\Lst})\bigl(f_0^{A^2}(m_{\Lambda_b}-m_{\Lst})^2 + 2f_\perp^{A^2}q^2\bigr)s^2_+s_-\biggr)\biggr]\, ,
\eea
where the LFV Wilson coefficients are 
\begin{align}\label{eq:LFV-WC}
\begin{split}
\mathbb{C}_-^{\rm LFV} &= \sum_{i\neq j}\bigg[\bigg|C^{ij}_{LL}-C^{ij}_{RL} \bigg|^2 + \bigg|C^{ij}_{LR}-C^{ij}_{RR} \bigg|^2 \bigg] \, ,\\
\mathbb{C}_+^{\rm LFV} &=  \sum_{i\neq j}\bigg[ \bigg|C^{ij}_{LL}+C^{ij}_{RL} \bigg|^2 + \bigg|C^{ij}_{LR}+C^{ij}_{RR} \bigg|^2\bigg]\, ,\\
\mathbb{C}^{\prime \rm LFV} &=  \sum_{i\neq j} \bigl(C^{ij}_{LL}-C^{ij}_{RL}\bigr)^\ast \bigl(C^{ij}_{LL}+C^{ij}_{RL}\bigr) + \sum_{i\neq j} \bigl(C^{ij}_{LR}-C^{ij}_{RR}\bigr)^\ast \bigl(C^{ij}_{LR}+C^{ij}_{RR}\bigr)\, .
\end{split}
\end{align}
Adding the SM differential branching ratios (which can be obtained from equations \eqref{eq:Lambda-dqdcos} and \eqref{eq:LamSt-dq} by taking NP Wilson coefficients to zero) to \eqref{eq:LFV-dBdq2-Lam} and \eqref{eq:LFV-dBdq2-LamSt}, the longitudinal polarization fraction and the hadronic side forward-backward asymmetry observales are extracted using \eqref{eq:FL-def} and \eqref{eq:AFBh-def}. The $F_L^{\Lambda}$ and the $A_{{\rm FB}}^{\Lambda}$ for the $\Lambda_b\to \Lambda(\to p\pi)\nu\bar{\nu}$ are

\bea\label{eq:FL-LFV}
F_L^\Lambda &=& \frac{1}{3}\,,\\
A_{\rm FB}^{h,\Lambda} &=& -\frac{\mathbb{H}_3\bigg(3|C^{\rm SM}|^2+\mathsf{Re}(\mathbb{C}^{\rm \prime LFV})\bigg)}{\mathbb{H}_1(3|C^{\rm SM}|^2+\mathbb{C}_+^{\rm LFV})+\mathbb{H}_2(3|C^{\rm SM}|^2+\mathbb{C}_-^{\rm LFV})}\, ,
\eea
where the $q^2$-dependent functions are 
\bea
\mathbb{H}_1 &=& \bigg(\bigl(m_{\Lambda_b}+m_{\Lambda}\bigr)^2s_{-}f_0^{V^2}+2q^2s_{-}f_{\perp}^{V^2} \bigg)\, ,\nn\\
\mathbb{H}_2 &=& \bigg(\bigl(m_{\Lambda_b}-m_{\Lambda}\bigr)^2s_{+}f_0^{A^2}+2q^2s_{+}f_{\perp}^{A^2} \bigg)\, ,\nn\\
\mathbb{H}_3 &=& -\alpha_\Lambda\bigg( 2 q^2 \sqrt{s_- s_+}f^A_{\perp}f^V_{\perp} +   \bigl(\mmLb -\mmL \bigr) \sqrt{s_- s_+}f^A_{0}f^V_{0} \bigg)
\, .\nn
\eea
For the $\Lambda_b\to \Lambda^\ast(\to N\bar{K})\nu\bar{\nu}$ mode the $F_{L}^{\Lambda^\ast}$ is  
\be 
F_{L}^{\Lst} = \frac{2\bigl[\mathbb{C}_+^{\rm LFV} \mathbb{G}_1+\mathbb{C}_-^{\rm LFV}\mathbb{G}_2+3 C^{\rm SM^2}(\mathbb{G}_1+\mathbb{G}_2)\bigr]}{3\bigl[\mathbb{C}_+^{\rm LFV}\bigl(\mathbb{G}_1^{\prime}+ \mathbb{G}_1 \bigr)+\mathbb{C}_-^{\rm LFV}\bigl(\mathbb{G}_2^{\prime}+ \mathbb{G}_2 \bigr)+3C^{\rm SM^2}(\mathbb{G}_1+\mathbb{G}_2+\mathbb{G}_1^{\prime}+\mathbb{G}_2^{\prime})\bigr]}\, ,
\ee

For numerical analysis the $q^2$-dependent form factors $f_0^{V,A}, f_\perp^{V,A}, f_g^{V,A}$  are taken from the lattice QCD calculations presented in \cite{Detmold:2016pkz,Meinel:2020owd,Meinel:2021mdj}. As elaborated in \cite{Meinel:2020owd} the $q^2$ parametrizations of the lattice calculations are not expected to be reliable below 16.3 GeV$^2$. Therefore, our numerical estimates of $\Lambda_b\to \Lambda^\ast(\to p\pi)\nu\bar{\nu}$ is restricted to $q^2 \geq 16.3$GeV$^2$. Apart from the form factor uncertainties we have included the uncertainties coming from the CKM matrix elements and the SM Wilson coefficient \eqref{eq:SMWC}. Integrating the differential branching ratios \eqref{eq:Lambda-dqdcos} and \eqref{eq:LamSt-dq} we present below the total (SM+NP) branching ratios when the NP is lepton flavor conserving  
\bea \label{eq:Int-Br}
\mathcal{B}_\Lambda &=& (7.84 \pm 0.94)\times 10^{-6}+(-2.34 \pm 0.35)\times 10^{-7}\sum_i \mathsf{Re}[(C^{ii}_{LL}-C^{ii}_{RL})^*]\, ,\nn\\ &-& (1.62 \pm 0.3)\times 10^{-7}\sum_i \mathsf{Re}[(C^{ii}_{LL}+C^{ii}_{RL})^*] \nn \\ &+& (8.90 \pm 1.33) \times 10^{-9} \sum_{i} \bigg[  |C_{LL}^{ii} - C_{RL}^{ii}|^2  +   |C_{LR}^{ii} - C_{RR}^{ii}|^2\bigg]\, ,\nn\\ &+& (6.15 \pm 1.10) \times 10^{-9} \sum_i\bigg[  |C_{LL}^{ii} + C_{RL}^{ii}|^2  +   |C_{LR}^{ii} + C_{RR}^{ii}|^2\bigg]\, ,\\
\mathcal{B}_{\Lst}  &=& (3.01 \pm 0.32)\times 10^{-9}+(-1.31 \pm 0.15)\times 10^{-10}\sum_i \mathsf{Re}[(C^{ii}_{LL}-C^{ii}_{RL})^*]\, ,\nn\\&-& (2.10 \pm 0.17)\times 10^{-11}\sum_i \mathsf{Re}[(C^{ii}_{LL}+C^{ii}_{RL})^*] \nn \\ &+&(4.99 \pm 0.59) \times 10^{-12}  \sum_{i} \bigg[  |C_{LL}^{ii} - C_{RL}^{ii}|^2  +   |C_{LR}^{ii} - C_{RR}^{ii}|^2\bigg]\, ,\nn\\&+& (7.97 \pm 0.64) \times 10^{-13} \sum_{i} \bigg[  |C_{LL}^{ii} + C_{RL}^{ii}|^2  +   |C_{LR}^{ii} + C_{RR}^{ii}|^2\bigg]\, .
\eea
When the NP is lepton flavor violating, the total (SM+NP) branching ratios are  
\bea \label{eq:Int-BrLFV}
\mathcal{B}_\Lambda &=& (6.09\pm 0.78)\times 10^{-6}+(8.90\pm 1.33)\times 10^{-9}~\mathbb{C}_-^{\rm LFV} + (6.15\pm 1.10)\times 10^{-9}~\mathbb{C}_+^{\rm LFV} \,, \\
\mathcal{B}_{\Lst}  &=&(3.01\pm 0.32)\times 10^{-9} + (4.99\pm 0.59)\times 10^{-12}\mathbb{C}_-^{\rm LFV} + (7.97\pm 0.64)\times 10^{-13}\mathbb{C}_+^{\rm LFV}\, .
\eea
In figure \ref{fig:LbL} we present the SM $q^2$ distribution of the $\Lambda_b\to\Lambda(\to p\pi)\nu\bar{\nu}$ and $\Lambda_b\to\Lambda^\ast(\to N\bar{K})\nu\bar{\nu}$ observables, respectively. The $q^2$ integrated results are presented in table \ref{tab:SM-pred}.

For the purpose of comparing future measurements with the SM predictions we also define the following ratios: in the case of lepton flavor conserving new physics  
\bea
R^{\Lambda}_{\nu\nu} &=& \frac{\mathcal{B}(\Lambda_b\to \Lambda \nu\bar{\nu})}{\mathcal{B}(\Lambda_b\to \Lambda \nu\bar{\nu})^{\rm SM}}\, ,\nn\\
&=& 1 +(-2.99\pm 0.45)\times 10^{-2}\mathsf{Re}(C_{LL}^{ii} - C_{RL}^{ii})^\ast+(-2.06\pm 0.37)\times 10^{-2}\mathsf{Re}(C_{LL}^{ii} + C_{RL}^{ii})^\ast \nn \\ &+& (1.13\pm 0.17)\times 10^{-3} \sum_{i} \bigg[  |C_{LL}^{ii} - C_{RL}^{ii}|^2  +   |C_{LR}^{ii} - C_{RR}^{ii}|^2 \bigg] \, \nn\\& + &
 (7.83\pm 1.40)\times 10^{-4} \sum_{i} \bigg[  |C_{LL}^{ii} + C_{RL}^{ii}|^2  +   |C_{LR}^{ii} + C_{RR}^{ii}|^2   \bigg]
\eea
\bea\label{eq:Rnunu-LamSt-16}
R^{\Lambda^
\ast}_{\nu\nu} &=& \frac{\mathcal{B}(\Lambda_b\to \Lambda^\ast \nu\bar{\nu})}{\mathcal{B}(\Lambda_b\to \Lambda^\ast \nu\bar{\nu})^{\rm SM}}\, ,\nn\\
&=& 1 +(-4.37\pm 0.51)\times 10^{-2}\mathsf{Re}(C_{LL}^{ii} - C_{RL}^{ii})^\ast+(-6.98\pm 0.56)\times 10^{-3}\mathsf{Re}(C_{LL}^{ii} + C_{RL}^{ii})^\ast \nn \\ &+& (1.66\pm 0.19)\times 10^{-3} \sum_{i} \bigg[  |C_{LL}^{ii} - C_{RL}^{ii}|^2  +   |C_{LR}^{ii} - C_{RR}^{ii}|^2 \bigg] \, \nn\\& + &
 (2.65\pm 0.21)\times 10^{-4} \sum_{i} \bigg[  |C_{LL}^{ii} + C_{RL}^{ii}|^2  +   |C_{LR}^{ii} + C_{RR}^{ii}|^2   \bigg]
\eea
In case of LFV new physics
\bea
R^\Lambda_{\nu\nu} &=& 1+\bigl(1.13\pm 0.17\bigr)\times 10^{-3}\sum_{i\neq j}\biggl[ \bigg|C^{ij}_{LL}-C^{ij}_{RL}\bigg|^2+\bigg|C^{ij}_{LR}-C^{ij}_{RR}\bigg|^2\biggr]\nn\\ &+& \bigl(7.83\pm 1.40\bigr)\times 10^{-4}\sum_{i\neq j}\biggl[ \bigg|C^{ij}_{LL}+C^{ij}_{RL}\bigg|^2+\bigg|C^{ij}_{LR}+C^{ij}_{RR}\bigg|^2\biggr] 
\eea
\bea
R^{\Lambda^\ast}_{\nu\nu} &=& 1+\bigl(4.97\pm 0.59\bigr)\times 10^{-3}\sum_{i\neq j}\biggl[ \bigg|C^{ij}_{LL}-C^{ij}_{RL}\bigg|^2+\bigg|C^{ij}_{LR}-C^{ij}_{RR}\bigg|^2\biggr]\nn\\ &+& \bigl(2.65\pm 0.21\bigr)\times 10^{-4}\sum_{i\neq j}\biggl[ \bigg|C^{ij}_{LL}+C^{ij}_{RL}\bigg|^2+\bigg|C^{ij}_{LR}+C^{ij}_{RR}\bigg|^2\biggr] 
\eea
\begin{figure}[ht]
\center 
\includegraphics[width=0.28\textwidth]{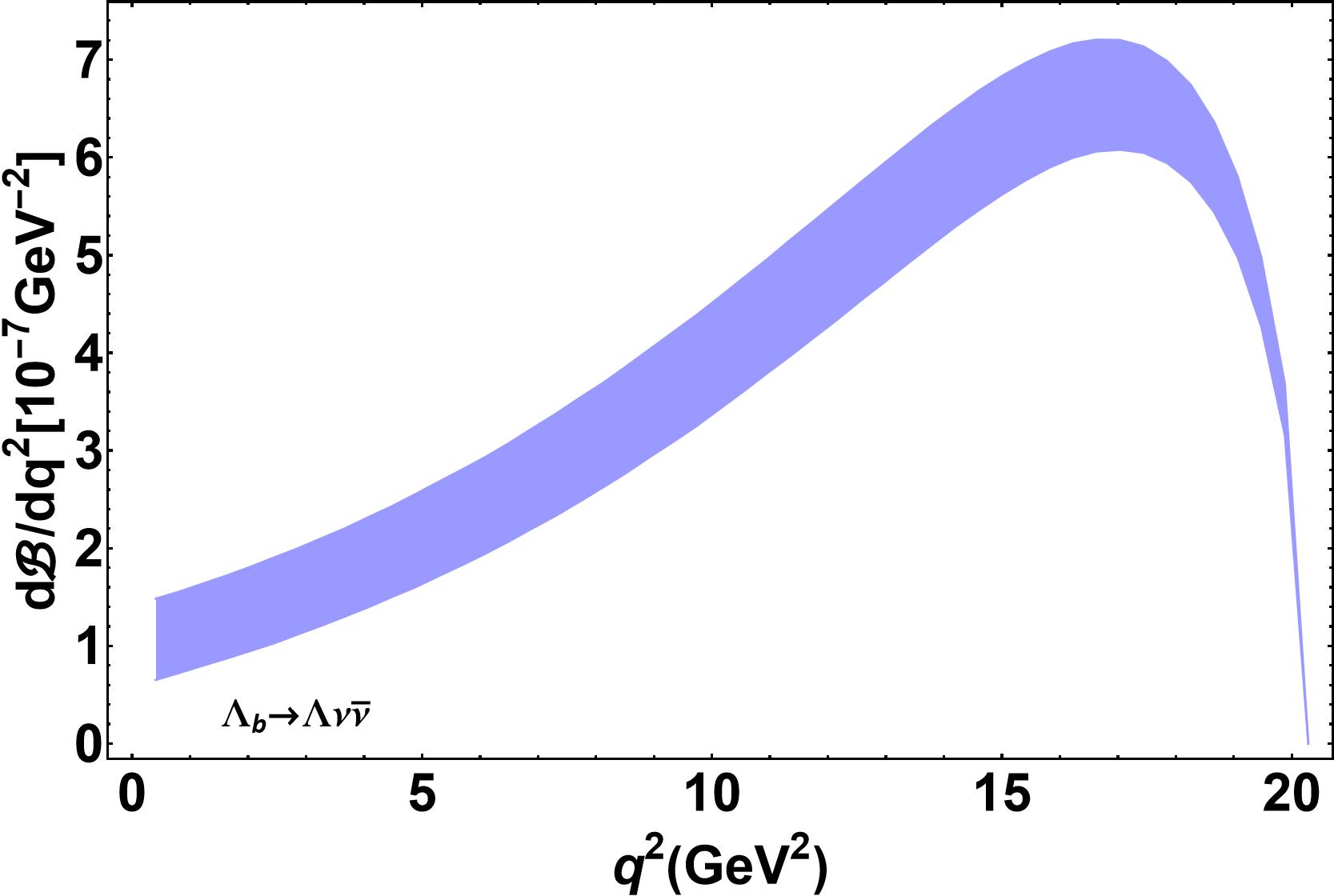}
\includegraphics[width=0.28\textwidth]{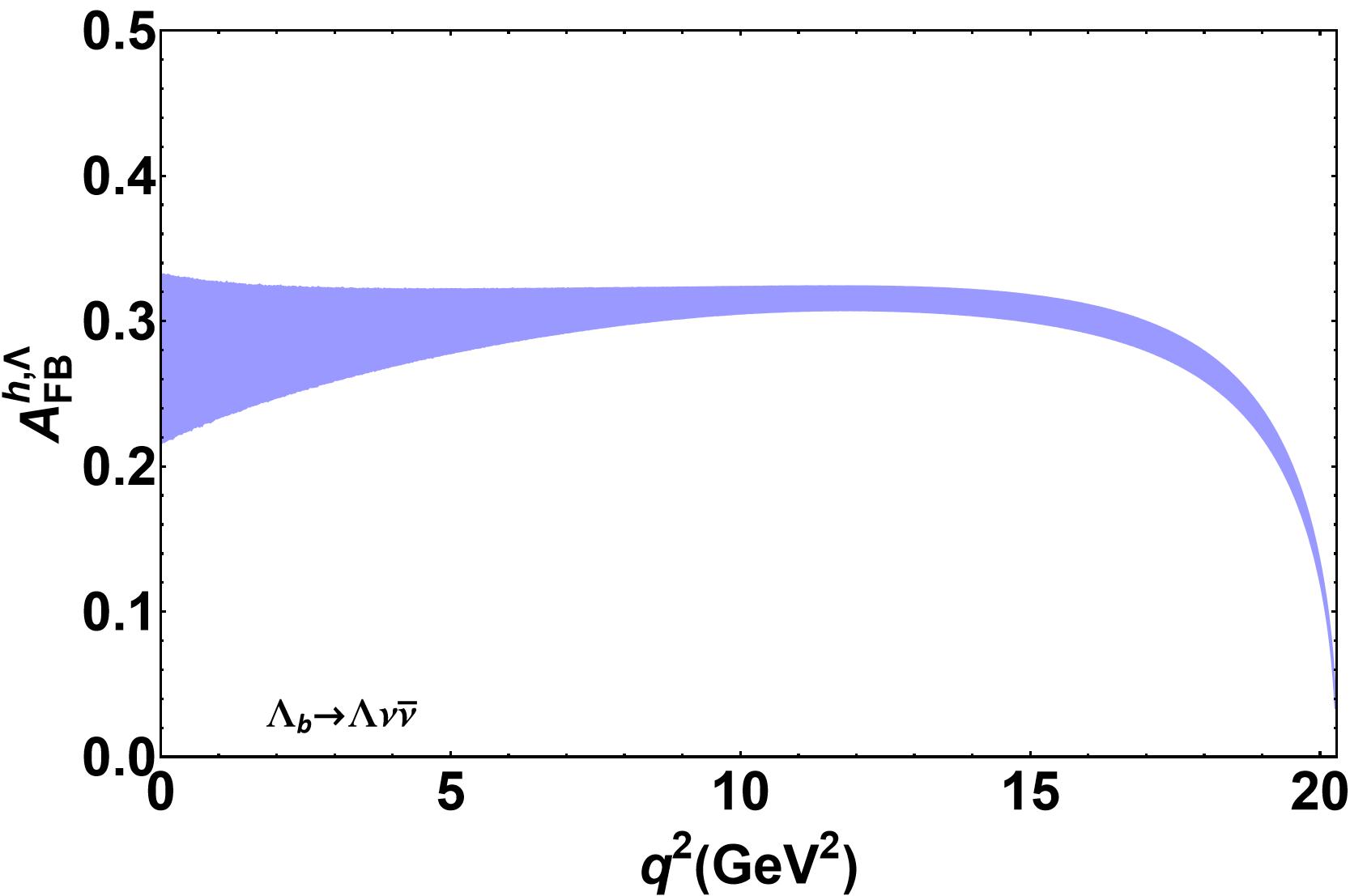}\\
\includegraphics[width=0.28\textwidth]{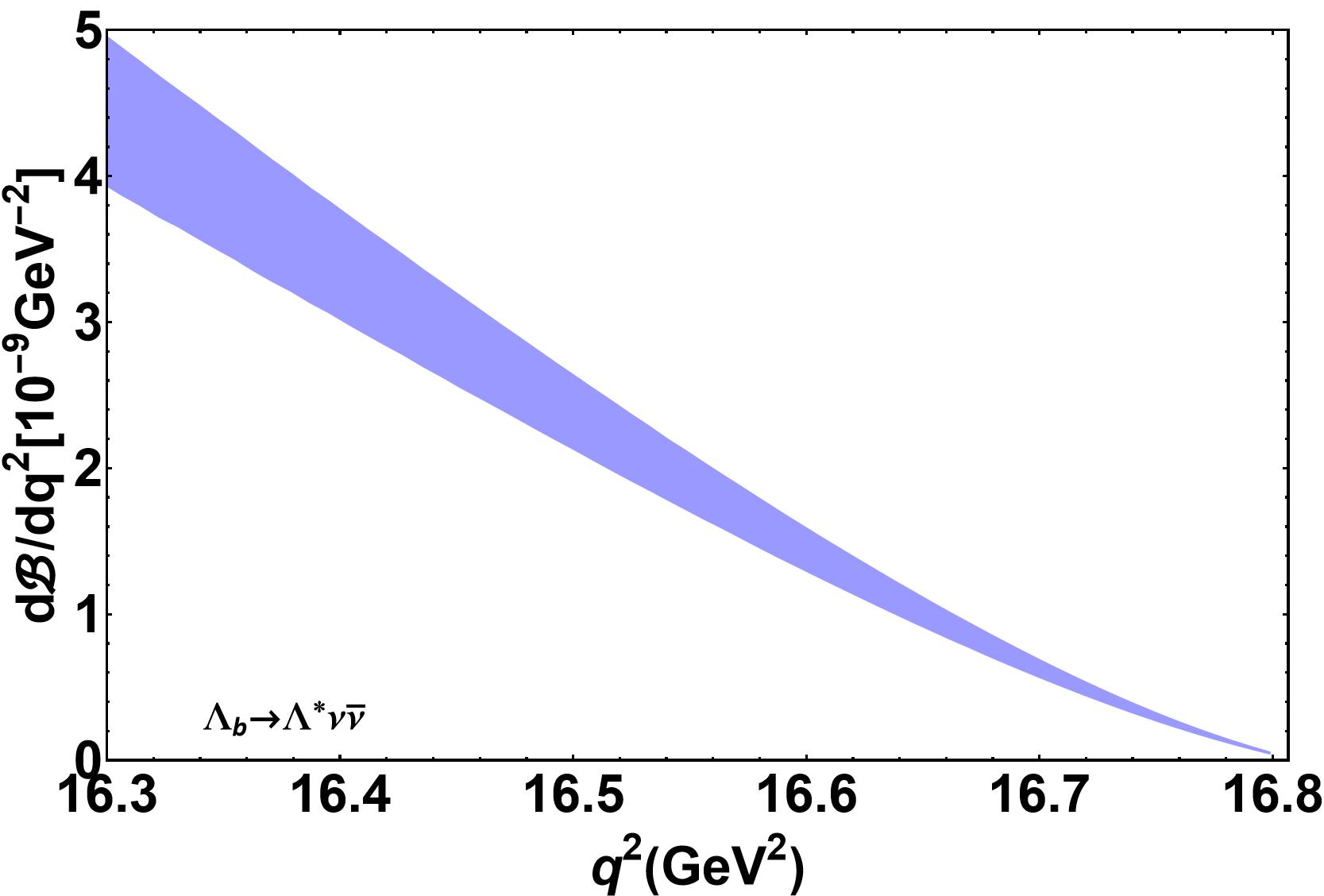}
\includegraphics[width=0.28\textwidth]{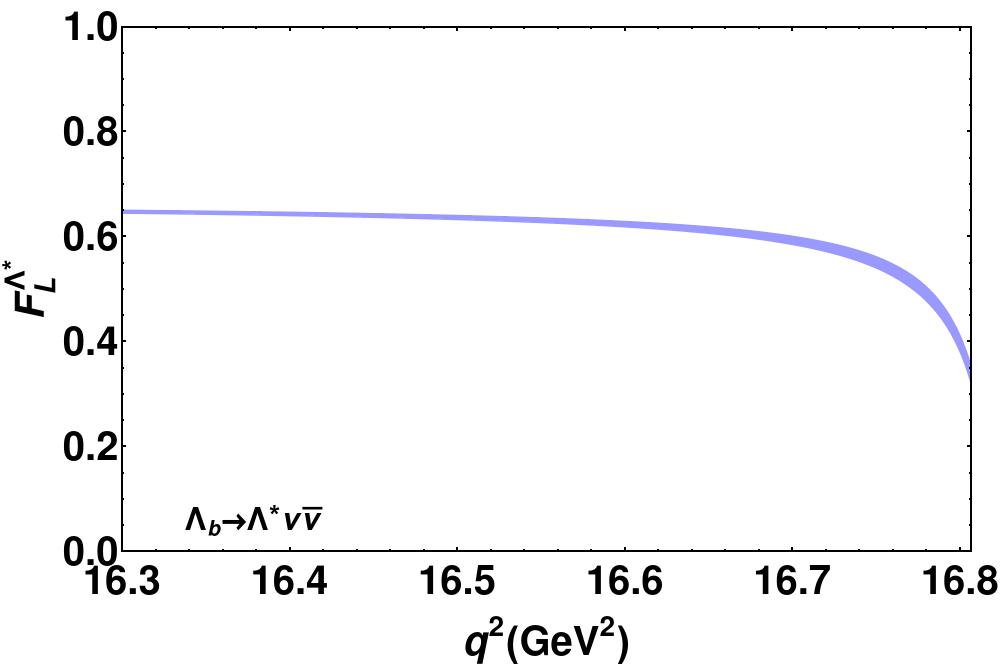}
\caption{ Standard Model prediction for $\Lambda_b\to \Lambda (\to p\pi) \bar{\nu} \nu$ and $\Lambda_b\to \Lambda^{\ast} (\to N\bar{K}) \bar{\nu} \nu$ observables.  The bands correspond to the uncertainties sourced by the form factors, the SM Wilson coefficients, and the CKM elements. }
\label{fig:LbL}
\end{figure}

\begin{table}[h!]
\begin{center}
\renewcommand{\arraystretch}{1.2} 
\begin{tabular}{c c c c}
\toprule\hline 
 & $\Lambda_b\to \Lambda (\to p\pi) \bar{\nu} \nu$ &  $\Lambda_b\to \Lambda^{\ast} (\to N\!\bar{K}) \bar{\nu} \nu$  \\
\midrule\hline 
$\mathcal{B}$ & $\bigl(7.84\pm 0.94\bigr)\times 10^{-6}$ &  $(3.01\pm 0.32)\times 10^{-9}$  \\
$A^{h}_{{\rm FB}}$& $0.27\pm0.03$ & $-$  \\
$F_{{\rm L}}$& $1/3$ & $0.635\pm 0.084$  \\
\toprule\hline 
\end{tabular}
\caption{Standard Model predictions for $\Lambda_b\to \Lambda (\to p\pi) \bar{\nu} \nu$ and $\Lambda_b\to \Lambda^{\ast} (\to N\bar{K}) \bar{\nu} \nu$ observables. The $\Lambda_b\to \Lambda^{\ast} (\to N\bar{K}) \bar{\nu} \nu$ observables are integrated from 16.3 GeV$^2$ to the end point due to the reason mentioned in the texts. } 
\label{tab:SM-pred}
\end{center}
\end{table}

\section{Implications of Belle-II \label{sec:belle-ii}}
To analyze the new physics implications of the Belle-II excess on the $\Lambda_b \to \Lambda^{(\ast)}\nu\bar{\nu}$ decays we utilize the the $R_K^{\nu\nu}$ and $R_{K^\ast}^{\nu\nu}$ data \eqref{eq:Belle-II-ratios}. The expressions of $R_K^{\nu\nu}$ and $R_{K^\ast}^{\nu\nu}$ following \cite{He:2021yoz, Browder:2021hbl}  are 
\begin{align}\label{eq:RKnunu}
R^{\nu\nu}_K &\approx 1-0.1~{\rm Re}\sum_i\left( C^{ii}_{LL}+C^{ii}_{RL}\right)
+0.008 \sum_{ij}\left( \left|C^{ij}_{LL}+C^{ij}_{RL}\right|^2 +\left|C_{LR}^{ij}+C_{RR}^{ij}\right|^2 \right)\,, \\
\label{eq:RKnunuSt}
R^{\nu\nu}_{K^*}&\approx  1+\sum_i{\rm Re}\left( -0.1~C^{ii}_{LL}+0.07~C^{ii}_{RL}\right) +  \sum_{ij}\bigg[0.008\left({|C^{ij}_{LL}}|^{2}+|{C^{ij}_{RL}}|^2 +|{C_{LR}^{ij}}|^2+|{C_{RR}^{ij}}|^2\right) \, \nn\\ &-0.01\left(C^{ij}_{LL}C^{ij}_{RL}+C_{LR}^{ij}C_{RR}^{ij}\right)\bigg ].
\end{align}
Following we discuss three scenarios: lepton flavor universal new physics, new physics with lepton flavor universality violation but conserved lepton number, and lepton flavor violating new physics.

{\it Lepton Flavor Universal new physics:} When the NP is lepton flavor universal, there is interference between the SM and the NP as seen in equation \eqref{eq:nonLFV}. The NP is described by the sets of new physics Wilson coefficients: $C^{ii}_{LL}$, $C^{ii}_{RL}$, $C^{ii}_{LR}$, and $C^{ii}_{RR}$. We consider the following two cases: in the first case both $C^{ii}_{LR}=0$ and $C^{ii}_{RR}=0$ for all flavors, and in the second case both $C^{ii}_{LL}=0$ and $C^{ii}_{RL}=0$ for all flavors. The upper limit on the branching ratios and upper and lower limits on $R^{\Lambda^{(\ast)}}_{\nu\nu}$ coming from the Belle-II data in the two cases are shown in tables \ref{tab:LFU-sc1} and \ref{tab:LFU-sc2}, respectively. In Table \ref{tab:LFU-sc1}, the $C^{ii}_{LL}$ and $C^{ii}_{RL}$ coefficients are varied for one lepton flavor, while the coefficient for the other two flavors were kept SM-like. In table \ref{tab:LFU-sc2} the $C^{ii}_{RR}$ and the $C^{ii}_{LR}$ were varied for one lepton flavor, while the two coefficients for the other flavors are SM like. 

In figure \ref{fig:LFU-NP}, we show the new physics sensitivity of $F_L^{\Lambda^\ast}$ and $A_{\rm FB}^{h,\Lambda}$. The first two plots from the left correspond to the case where $C^{ii}_{LL}=0$ and $C^{ii}_{RL}=0$, while the last two plots correspond to the case where $C^{ii}_{LR}=0$ and $C^{ii}_{RR}=0$. In each plots the set of non-zero coefficients was randomly varied subject to $R_{K^{\ast}}^{\nu\nu}$ constraints. In all the plots, in addition to the SM shown in black, the other colors indicate different NP scenarios. 
\begin{figure}[ht!]
\center 
\includegraphics[width=0.3\textwidth]{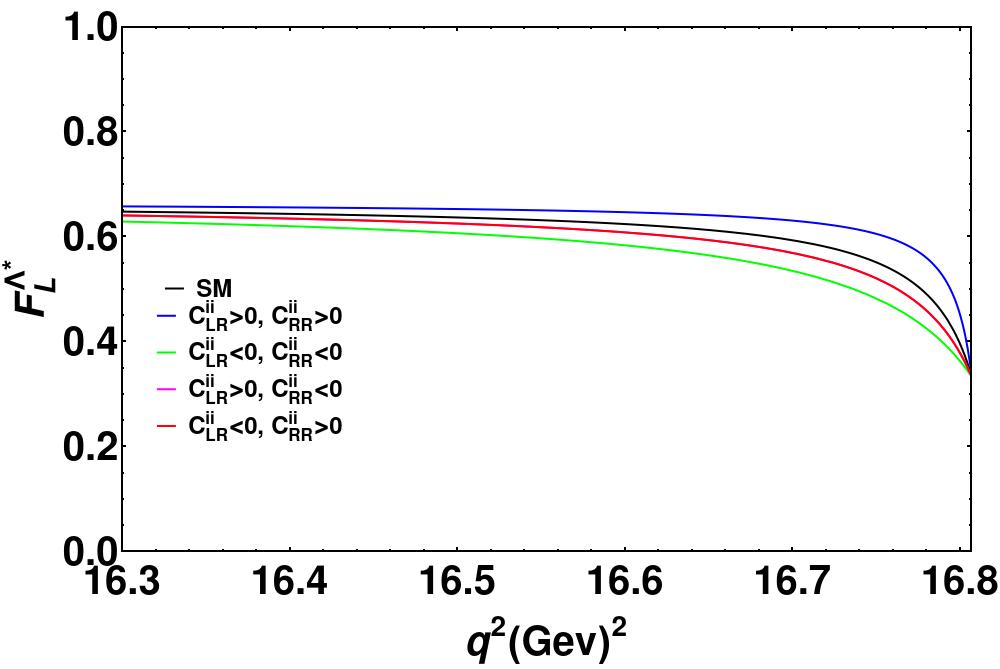}
\includegraphics[width=0.3\textwidth]{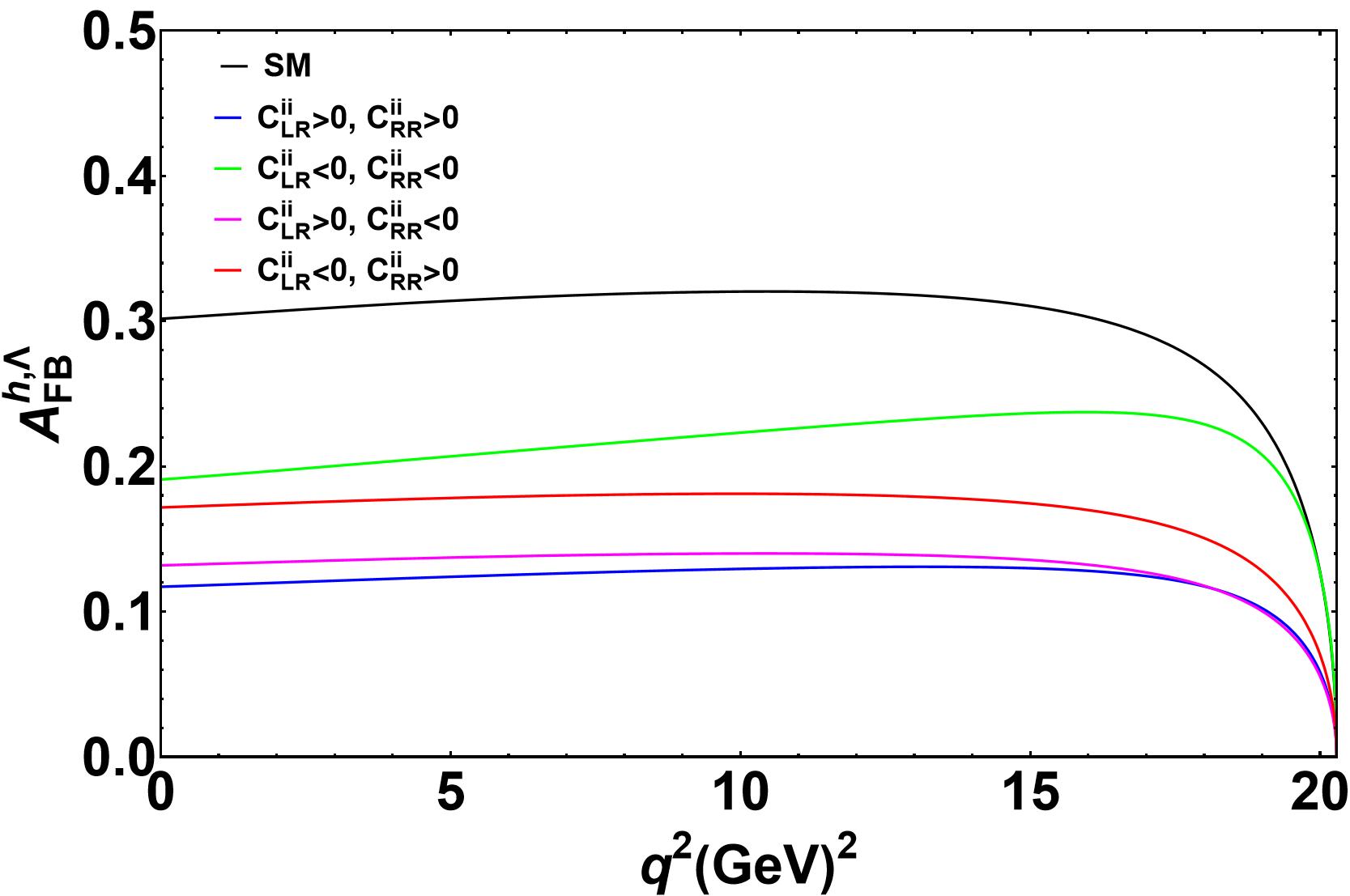}\\
\includegraphics[width=0.3\textwidth]{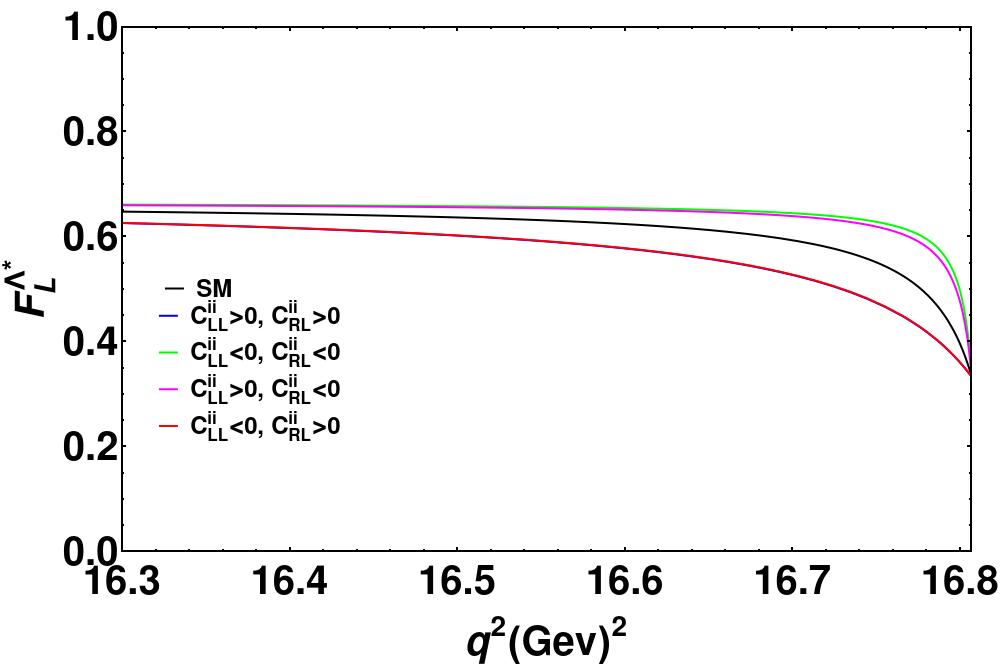} 
\includegraphics[width=0.3\textwidth]{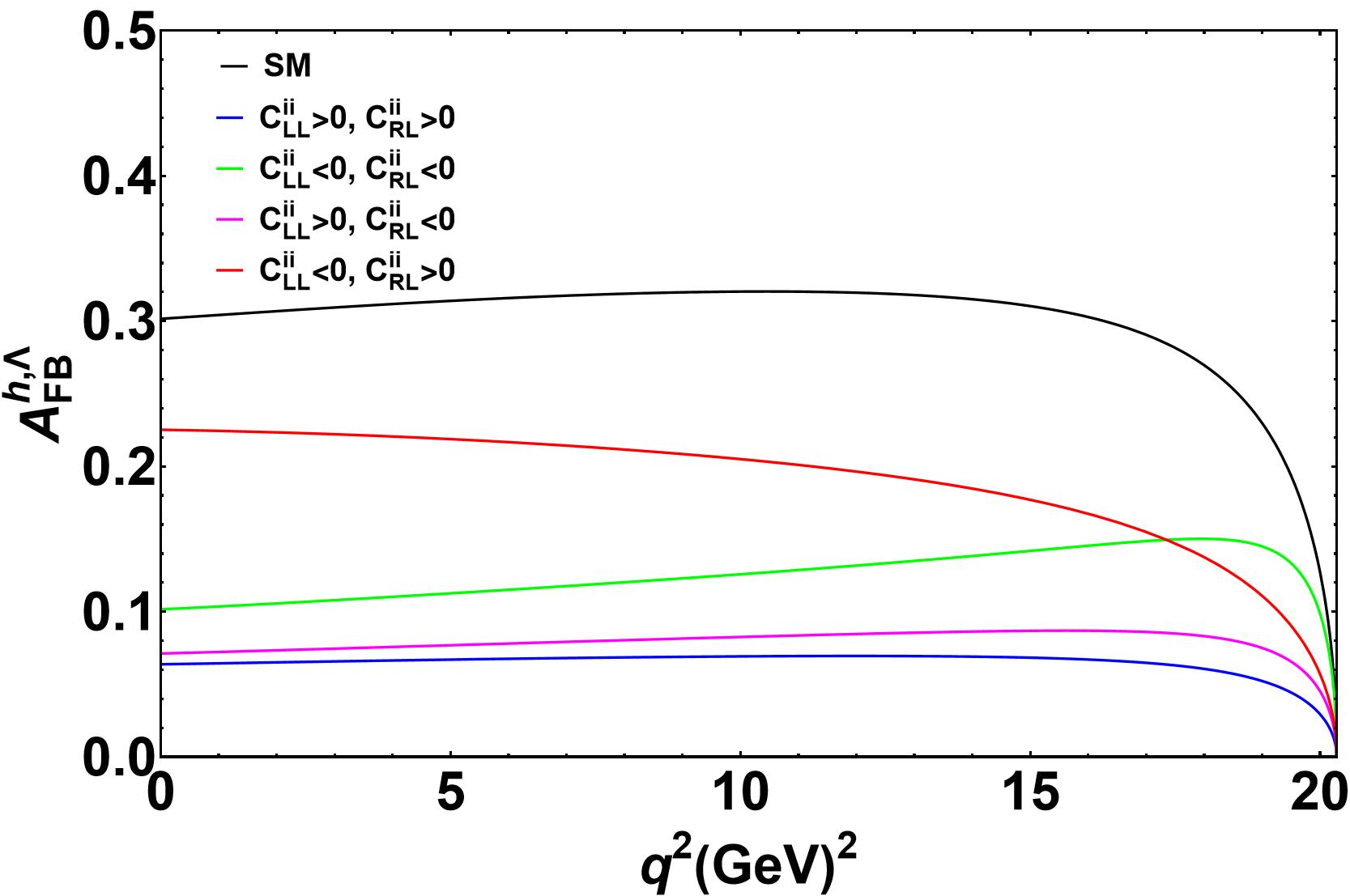}
\caption{The observables $F_L^{\Lambda^\ast}$ and $A_{\rm FB}^{h,\Lambda}$ in the scenario with lepton flavor universal new physics. The two plots on the top correspond to the case where $C^{ii}_{LL}=0$, $C^{ii}_{RL}=0$. The two plots on the bottom correspond to case where $C^{ii}_{LR}=0$ and $C^{ii}_{RR}=0$.}
\label{fig:LFU-NP}
\end{figure}

\begin{table*}[ht]
\begin{center}
\renewcommand{\arraystretch}{0.1} 
\begin{tabular}{c c c c c}
\toprule \hline
\vspace{0.0001mm}\\
& \parbox[c]{2.5cm}{$C^{ii}_{LL}>0$ \\ $C^{ii}_{RL}>0$} &  \parbox[c]{2.5cm}{$C^{ii}_{LL}<0$ \\ $C^{ii}_{RL}<0$} & \parbox[c]{2.5cm}{$C^{ii}_{LL}>0$ \\ $C^{ii}_{RL}<0$} & \parbox[c]{2.5cm}{$C^{ii}_{LL}<0$ \\ $C^{ii}_{RL}>0$} \\
\vspace{0.001mm}\\
\midrule\hline 
\vspace{0.01mm}\\
$\mathcal{B}_\Lambda$ & \parbox[c]{2.5cm}{$8.68\times 10^{-6}$}  & \parbox[c]{2.5cm} {$1.09\times 10^{-5}$} &  \parbox[c]{2.5cm}{$8.14\times 10^{-6}$} & \parbox[c]{2.5cm}{$1.00\times 10^{-5}$} \\
\vspace{0.01mm}\\
${R}_\Lambda$  & \parbox[c]{2.5cm}{$1.10$ \\ $0.73$}  & \parbox[c]{2.5cm} {$1.39$ \\ $0.99$} &  \parbox[c]{2.5cm}{$1.04$ \\ $0.73$} & \parbox[c]{2.5cm}{$1.27$ \\ $1.00$} \\
\vspace{0.01mm}\\
$\mathcal{B}_{\Lambda^{\ast}}$ & \parbox[c]{2.5cm}{$1.06\times 10^{-8}$ }  & \parbox[c]{2.5cm} {$1.08\times 10^{-8}$ } &  \parbox[c]{2.5cm}{$8.44\times 10^{-9}$ } & \parbox[c]{2.5cm}{$1.08\times 10^{-8}$ } \\
\vspace{0.01mm}\\
${R}_{\Lambda^{\ast}}$ & \parbox[c]{2.5cm}{$1.25$ \\ $0.73$}  & \parbox[c]{2.5cm} {$1.28$ \\ $0.83$} &  \parbox[c]{2.5cm}{$1.0$ \\ $0.69$} & \parbox[c]{2.5cm}{$1.28$ \\ $1.00$} \\
\vspace{0.01mm}\\
\toprule\hline 
\end{tabular}
\caption{Bounds on the $\Lambda_b\to \Lambda (\to p\pi) \bar{\nu} \nu$ and $\Lambda_b\to \Lambda^{\ast} (\to N\bar{K}) \bar{\nu} \nu$ observables in different lepton flavor universal NP scenarios. Keeping $C^{ii}_{RR}=0$, $C^{ii}_{LR}=0$ for all three flavors $i=e,\mu,\tau$, the $C^{ii}_{LL}$, $C^{ii}_{RL}$ are varied for one lepton flavor $i$ subject to the Belle-II constraints.}  
\label{tab:LFU-sc1}
\end{center}
\end{table*}

\begin{table*}[ht]
\begin{center}
\renewcommand{\arraystretch}{0.1} 
\begin{tabular}{c c c }
\toprule \hline
\vspace{0.0001mm}\\
& \parbox[c]{2.5cm}{$C^{ii}_{RR}>0$ \\ $C^{ii}_{LR}>0$} &  \parbox[c]{2.5cm}{$C^{ii}_{RR}>0$ \\ $C^{ii}_{LR}<0$}  \\
\vspace{0.001mm}\\
\midrule\hline 
\vspace{0.01mm}\\
$\mathcal{B}_\Lambda$ & \parbox[c]{2.5cm}{$8.98\times 10^{-6}$ }  &   \parbox[c]{2.5cm}{$8.61\times 10^{-6}$ }  \\
\vspace{0.01mm}\\
${R}_\Lambda$  & \parbox[c]{2.5cm}{$1.14$ \\ $1.00$}  &   \parbox[c]{2.5cm}{$1.1$ \\ $1.00$}  \\
\vspace{0.01mm}\\
$\mathcal{B}_{\Lambda^{\ast}}$ & \parbox[c]{2.5cm}{$9.27\times 10^{-9}$ }  &  \parbox[c]{2.5cm}{$9.32\times 10^{-9}$ }  \\
\vspace{0.01mm}\\
${R}_{\Lambda^{\ast}}$ & \parbox[c]{2.5cm}{$1.10$ \\ $1.00$}  &   \parbox[c]{2.5cm}{$1.10$ \\ $1.00$}  \\
\vspace{0.01mm}\\
\toprule\hline 
\end{tabular}
\caption{Same as in table \ref{tab:LFU-sc1} but $C^{ii}_{LL}=0$, $C^{ii}_{RL}=0$ for all choices of $i$, and $C^{ii}_{RR}$, $C^{ii}_{LR}$ for a single choice of $i$ are varied subject to the Belle-II constraints.  As can be seen fro  \eqref{eq:nonLFV}, the case $C_{RR}^{ii}<0, C_{LR}^{ii}<0$ is identical to the case $C_{RR}^{ii}>0, C_{LR}^{ii}>0$, wheile $C_{RR}^{ii}<0$, $C_{LR}^{ii}>0$ is identical to the case $C_{RR}^{ii}>0$, $C_{LR}^{ii}<0$.} 
\label{tab:LFU-sc2}
\end{center}
\end{table*}

\begin{table*}[ht]
\begin{center}
\renewcommand{\arraystretch}{0.1} 
\begin{tabular}{c c c c c}
\toprule \hline
\vspace{0.0001mm}\\
& \parbox[c]{2.5cm}{$C^{ii}_{LL}>0$ \\ $C^{ii}_{RL}>0$} &  \parbox[c]{2.5cm}{$C^{ii}_{LL}<0$ \\ $C^{ii}_{RL}<0$} & \parbox[c]{2.5cm}{$C^{ii}_{LL}>0$ \\ $C^{ii}_{RL}<0$} & \parbox[c]{2.5cm}{$C^{ii}_{LL}<0$ \\ $C^{ii}_{RL}>0$} \\
\vspace{0.001mm}\\
\midrule\hline 
\vspace{0.01mm}\\
$\mathcal{B}_\Lambda$ & \parbox[c]{2.5cm}{$8.66\times 10^{-6}$ }  & \parbox[c]{2.5cm} {$1.13\times 10^{-5}$ } &  \parbox[c]{2.5cm}{$8.23\times 10^{-6}$ } & \parbox[c]{2.5cm}{$1.02\times 10^{-5}$ } \\
\vspace{0.01mm}\\
${R}_\Lambda$  & \parbox[c]{2.5cm}{$1.10$ \\ $0.46$}  & \parbox[c]{2.5cm} {$1.44$ \\ $0.98$} &  \parbox[c]{2.5cm}{$1.05$ \\ $0.46$} & \parbox[c]{2.5cm}{$1.30$ \\ $1.02$} \\
\vspace{0.01mm}\\
$\mathcal{B}_{\Lambda^{\ast}}$ & \parbox[c]{2.5cm}{$1.09\times 10^{-8}$ }  & \parbox[c]{2.5cm} {$1.41\times 10^{-8}$ } &  \parbox[c]{2.5cm}{$1.42\times 10^{-8}$ } & \parbox[c]{2.5cm}{$1.08\times 10^{-8}$ } \\
\vspace{0.01mm}\\
${R}_{\Lambda^{\ast}}$ & \parbox[c]{2.5cm}{$1.29$ \\ $0.47$}  & \parbox[c]{2.5cm} {$1.67$ \\ $0.75$} &  \parbox[c]{2.5cm}{$1.68$ \\ $0.47$} & \parbox[c]{2.5cm}{$1.28$ \\ $0.76$} \\
\vspace{0.01mm}\\
\toprule\hline 
\end{tabular}
\caption{Bounds on the $\Lambda_b\to \Lambda (\to p\pi) \bar{\nu} \nu$ and $\Lambda_b\to \Lambda^{\ast} (\to N\bar{K}) \bar{\nu} \nu$ observables in different lepton flavor universality violating NP scenarios. Keeping $C^{ii}_{RR}=0$, $C^{ii}_{LR}=0$ for all three flavors $i=e,\mu,\tau$, while the $C^{ii}_{LL}$, $C^{ii}_{RL}$ are varied for two lepton flavors subject to the Belle-II constraints.} 
\label{tab:LFUV-sc1}
\end{center}
\end{table*}

\begin{table*}[ht]
\begin{center}
\renewcommand{\arraystretch}{0.1} 
\begin{tabular}{c c c}
\toprule \hline
\vspace{0.0001mm}\\
& \parbox[c]{2.5cm}{$C^{ii}_{RR}>0$ \\ $C^{ii}_{LR}>0$} &  \parbox[c]{2.5cm}{$C^{ii}_{RR}>0$ \\ $C^{ii}_{LR}<0$}  \\
\vspace{0.001mm}\\
\midrule\hline 
\vspace{0.01mm}\\
$\mathcal{B}_\Lambda$ & \parbox[c]{2.5cm}{$9.15\times 10^{-6}$ }  &   \parbox[c]{2.5cm}{$8.60\times 10^{-6}$ }  \\
\vspace{0.01mm}\\
${R}_\Lambda$  & \parbox[c]{2.5cm}{$1.17$ \\ $1.00$}  &   \parbox[c]{2.5cm}{$1.1$ \\ $1.00$}  \\
\vspace{0.01mm}\\
$\mathcal{B}_{\Lambda^{\ast}}$ & \parbox[c]{2.5cm}{$9.27\times 10^{-9}$ }  &   \parbox[c]{2.5cm}{$9.25\times 10^{-9}$ }  \\
\vspace{0.01mm}\\
${R}_{\Lambda^{\ast}}$ & \parbox[c]{2.5cm}{$1.10$ \\ $1.00$}  &   \parbox[c]{2.5cm}{$1.10$ \\ $1.00$}  \\
\vspace{0.01mm}\\
\toprule\hline 
\end{tabular}
\caption{Same as in table \ref{tab:LFUV-sc1} but $C^{ii}_{LL}=0$, $C^{ii}_{RL}=0$ for all choices of $i$, and $C^{ii}_{RR}$, $C^{ii}_{LR}$ for a two choices of $i$ are varied subject to the Belle-II constraints. As can be seen from  \eqref{eq:nonLFV}, the case $C_{RR}^{ii}<0, C_{LR}^{ii}<0$ is identical to the case $C_{RR}^{ii}>0, C_{LR}^{ii}>0$, wheile $C_{RR}^{ii}<0$, $C_{LR}^{ii}>0$ is identical to the case $C_{RR}^{ii}>0$, $C_{LR}^{ii}<0$. } 
\label{tab:LFUV-sc2}
\end{center}
\end{table*}

{\it{Lepton Flavor Universality violating new physics:}} Here we consider a scenario in which the lepton flavor universality is violated but the lepton flavor is still conserved. The analysis is performed in the same way as in the previous scenario. We consider the following two cases: in the first case both $C^{ii}_{LR}=0$ and $C^{ii}_{RR}=0$ for all flavors, and in the second case both $C^{ii}_{LL}=0$ and $C^{ii}_{RL}=0$ for all flavors. The upper limit on branching ratios and upper and lower limits on $R^{\Lambda^{(\ast)}}_{\nu\nu}$ coming from the Belle-II data in different cases is shown in tables \ref{tab:LFUV-sc1} and \ref{tab:LFUV-sc2}, respectively. In table \ref{tab:LFUV-sc1}, $C^{ii}_{LL}$ and $C^{ii}_{RL}$ are varied for two lepton flavors independently while the coefficients for the other flavor is SM-like. In table \ref{tab:LFUV-sc2} the $C^{ii}_{RR}$ and $C^{ii}_{LR}$ are varied for two lepton flavors independently while the coefficients for the other flavor is SM like. To prepare tables \ref{tab:LFUV-sc1} and \ref{tab:LFUV-sc2}, the uncertainties in the theoretical inputs have been included. 

 In figure \ref{fig:LFUV-NP} we show the new physics sensitivity of $F_L^{\Lambda^\ast}$ and $A_{\rm FB}^{h,\Lambda}$. The first two plots from left correspond to scenario in which $C^{ii}_{LL}=0$, $C^{ii}_{RL}=0$ and the last two plots correspond to $C^{ii}_{LR}=0$ and $C^{ii}_{RR}=0$. In each plot the set of non-zero coefficients were randomly varied subject to the $R_{K^{\ast}}^{\nu\nu}$ constraints. In all the plots, in addition to the SM shown in black, the other colors indicate different NP scenarios. The benchmark points have been chosen for better distinguishability with the SM line. 
\begin{figure}[ht!]
\center 
\includegraphics[width=0.3\textwidth]{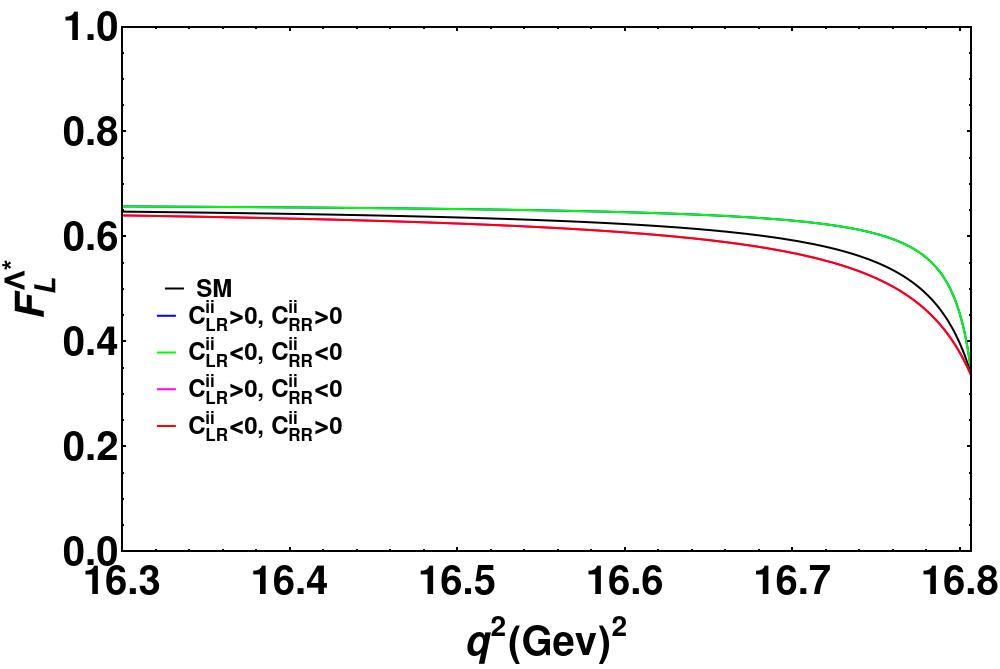}
\includegraphics[width=0.3\textwidth]{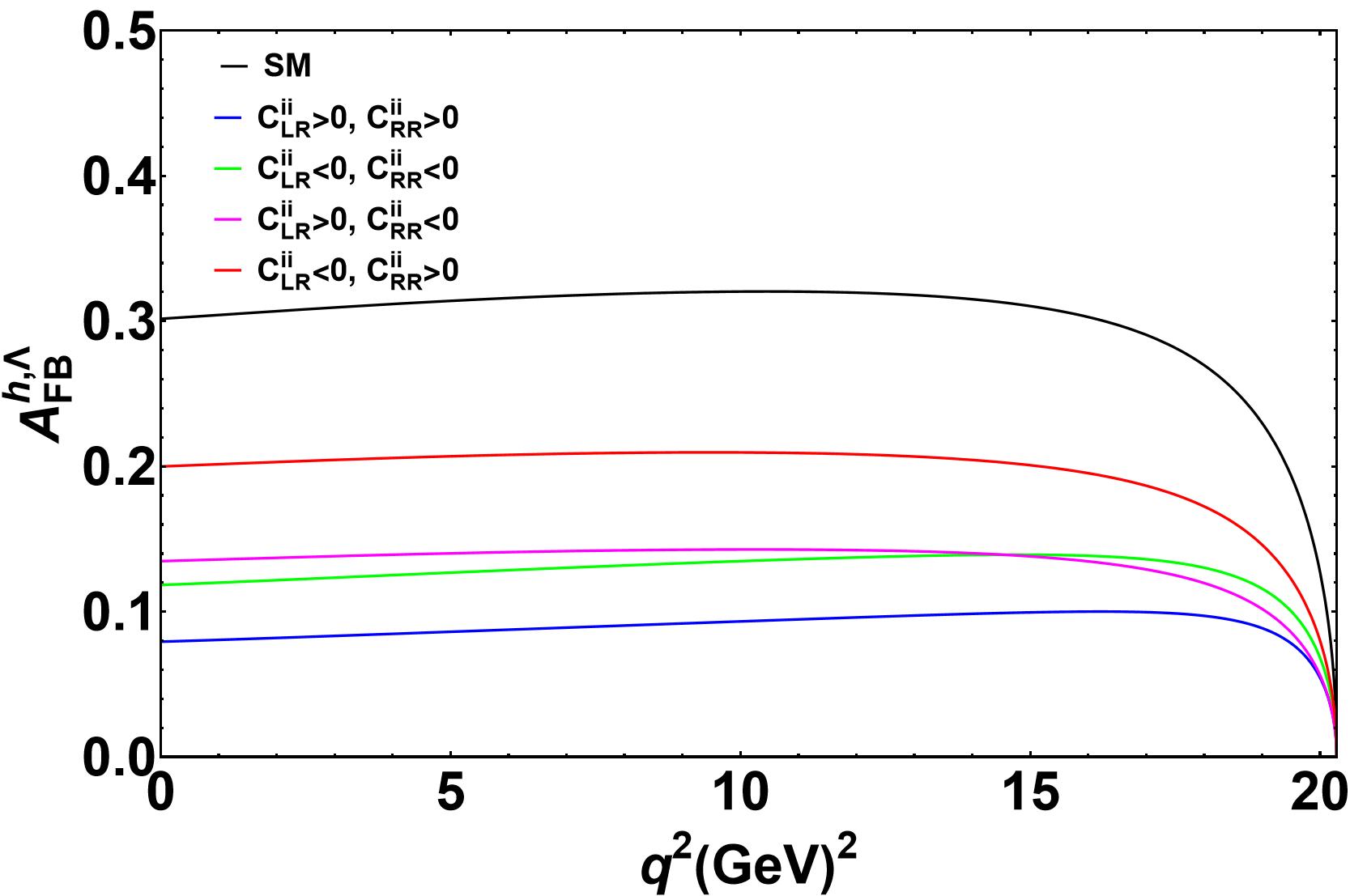}\\
\includegraphics[width=0.3\textwidth]{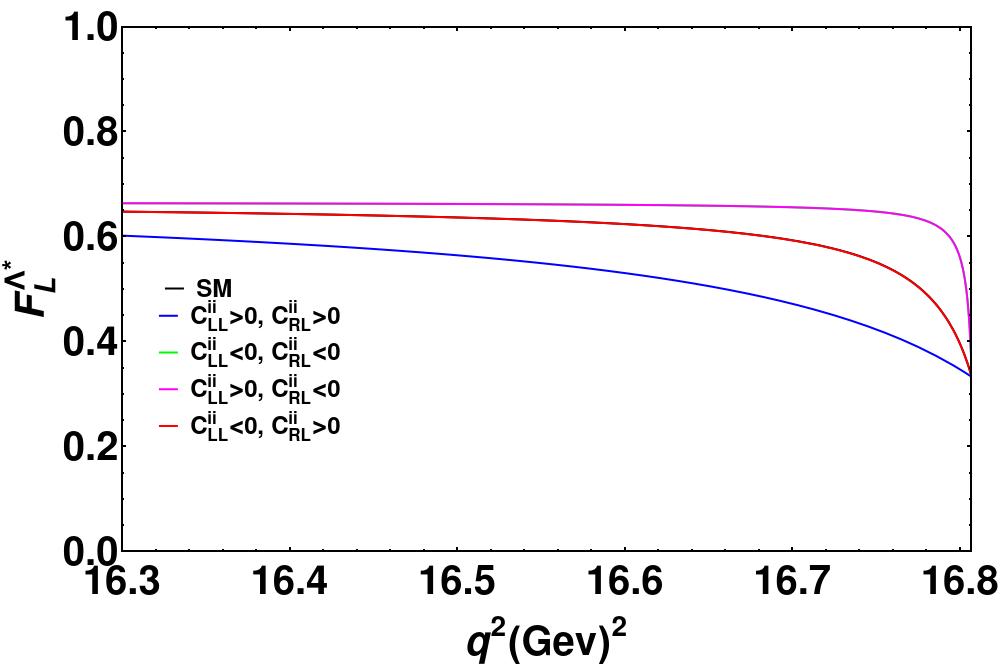} 
\includegraphics[width=0.3\textwidth]{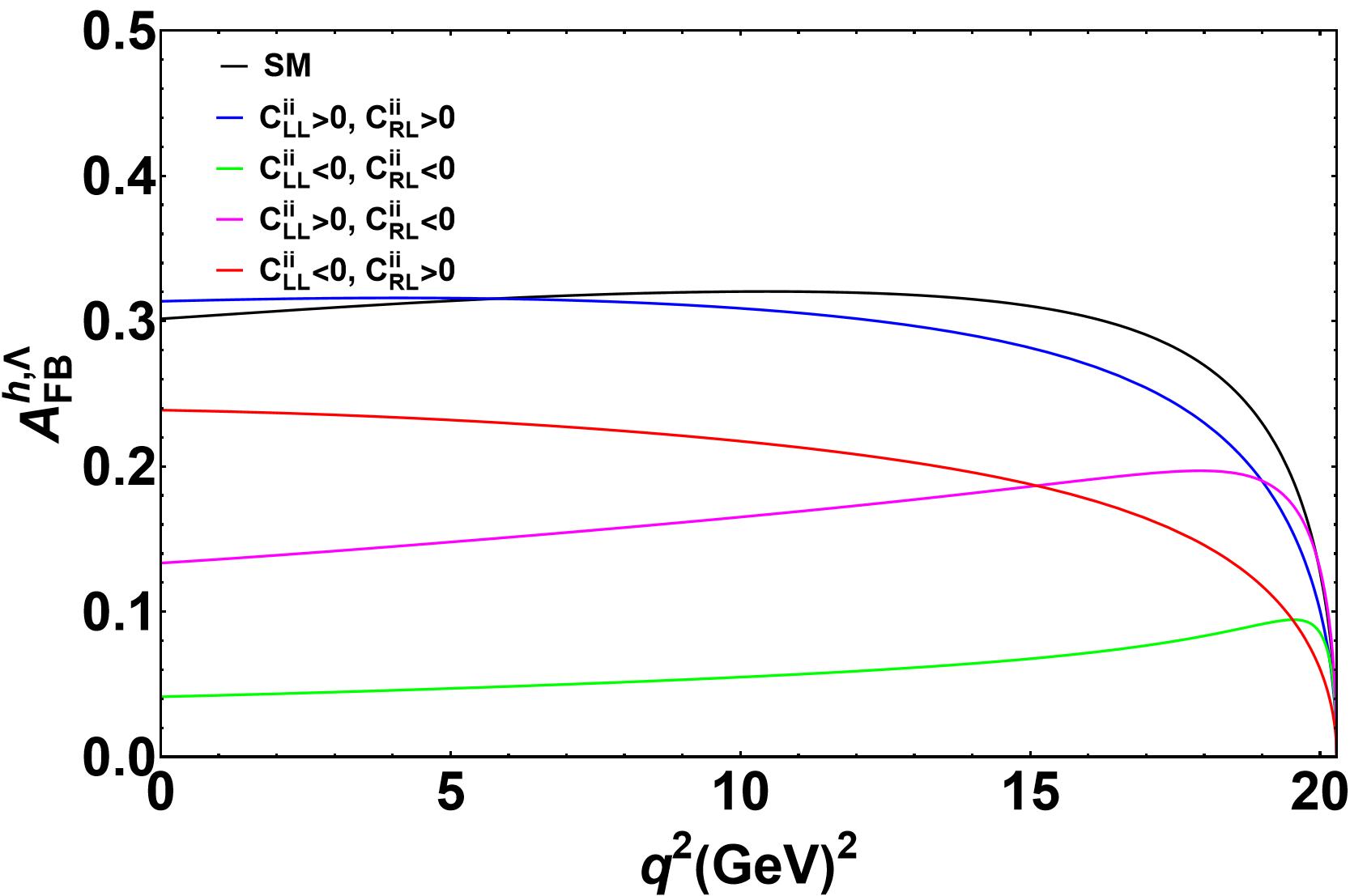}
\caption{The observables $F_L^{\Lambda^\ast}$ and $A_{\rm FB}^{h,\Lambda}$ in the scenario with lepton flavor universality violation but conserved lepton number. The first two plots from the left correspond to scenario where $C^{ii}_{LL}=0$, $C^{ii}_{RL}=0$. The last two plots from the left correspond to scenario where $C^{ii}_{LR}=0$ and $C^{ii}_{RR}=0$.}
\label{fig:LFUV-NP}
\end{figure}

\begin{table*}[ht]
\begin{center}
\renewcommand{\arraystretch}{0.1} 
\begin{tabular}{c c c}
\toprule \hline
\vspace{0.0001mm}\\
& \parbox[c]{2.5cm}{$C^{ii}_{LL}>0$ \\ $C^{ii}_{RL}>0$} &  \parbox[c]{2.5cm}{$C^{ii}_{LL}>0$ \\ $C^{ii}_{RL}<0$}  \\
\vspace{0.001mm}\\
\midrule\hline 
\vspace{0.01mm}\\
$\mathcal{B}_\Lambda$ & \parbox[c]{2.5cm}{$7.38\times 10^{-6}$ }  & \parbox[c]{2.5cm}{$6.83\times 10^{-6}$ }  \\
\vspace{0.01mm}\\
${R}_\Lambda$  & \parbox[c]{2.5cm}{$1.17$ \\ $1.00$}  &  \parbox[c]{2.5cm}{$1.09$ \\ $1.00$}  \\
\vspace{0.01mm}\\
$\mathcal{B}_{\Lambda^{\ast}}$ & \parbox[c]{2.5cm}{$3.30\times 10^{-9}$ }  & \parbox[c]{2.5cm}{$3.32\times 10^{-9}$ } \\
\vspace{0.01mm}\\
${R}_{\Lambda^{\ast}}$ & \parbox[c]{2.5cm}{$1.26$ \\ $1.00$}  &   \parbox[c]{2.5cm}{$1.25$ \\ $1.00$}  \\
\vspace{0.01mm}\\
\toprule\hline 
\end{tabular}
\caption{Upper bounds on the $\Lambda_b\to \Lambda (\to p\pi) \bar{\nu} \nu$ and $\Lambda_b\to \Lambda^{\ast} (\to N\bar{K}) \bar{\nu} \nu$ observables in different lepton flavor violating NP scenarios. Keeping $C^{ii}_{RR}=0$,$C^{ii}_{LR}=0$ for all three flavors $i=e,\mu,\tau$, the $C^{ii}_{LL}$,$C^{ii}_{RL}$ are varied for three lepton flavor $i$ subject to the Belle-II constraints. } 
\label{tab:LFV-sc2}
\end{center}
\end{table*}

\begin{figure}[h!]
\center 
\includegraphics[width=0.3\textwidth]{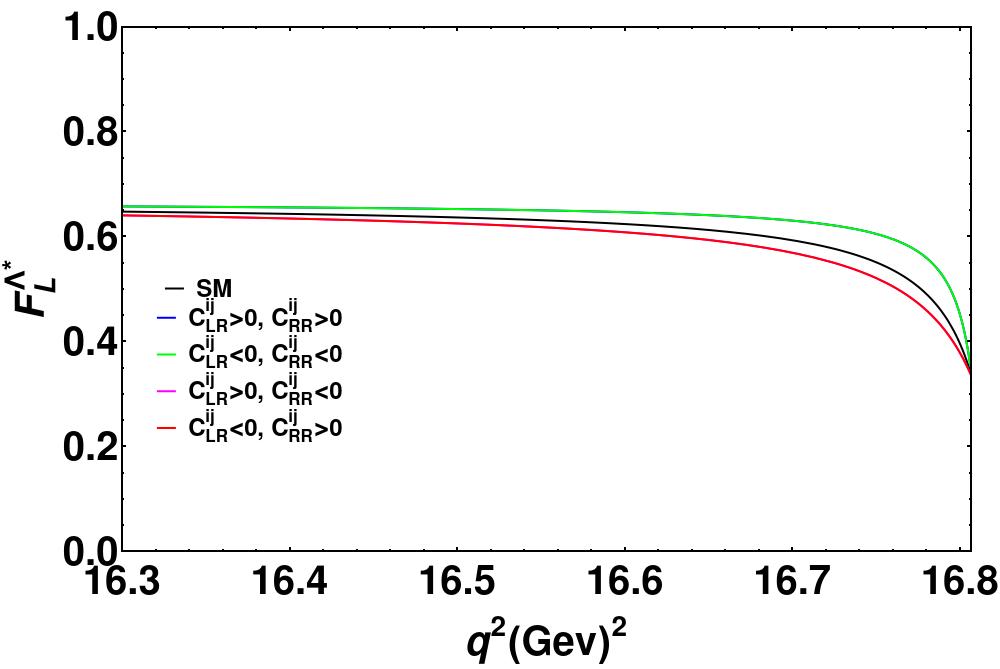} 
\includegraphics[width=0.3\textwidth]{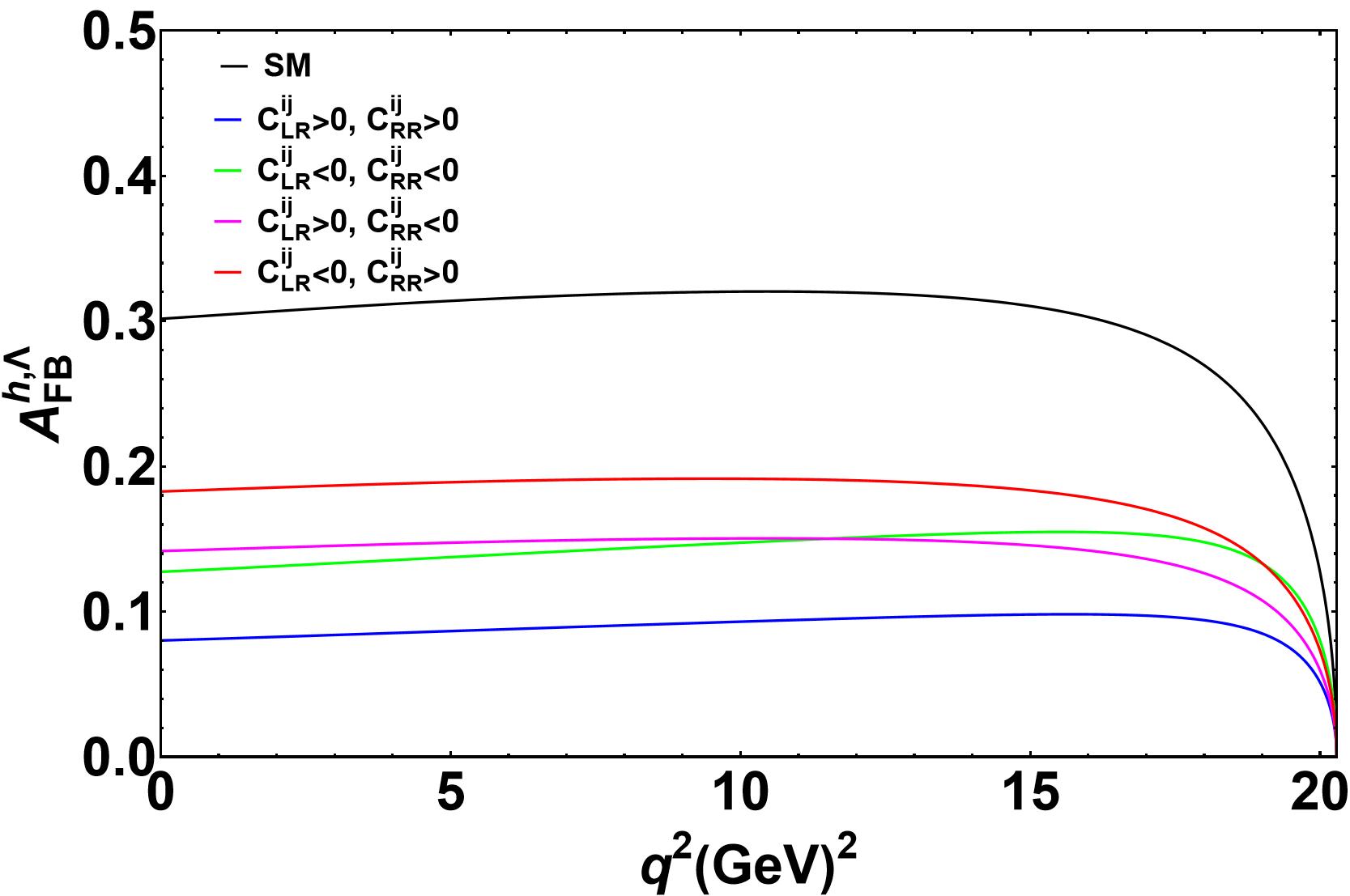}\\
\includegraphics[width=0.3\textwidth]{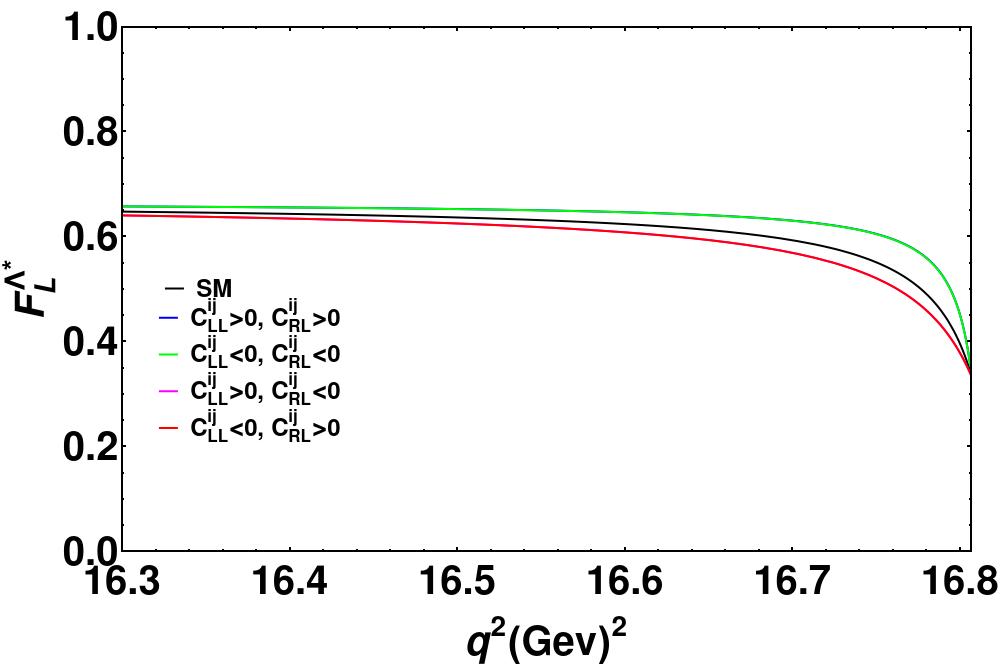} 
\includegraphics[width=0.3\textwidth]{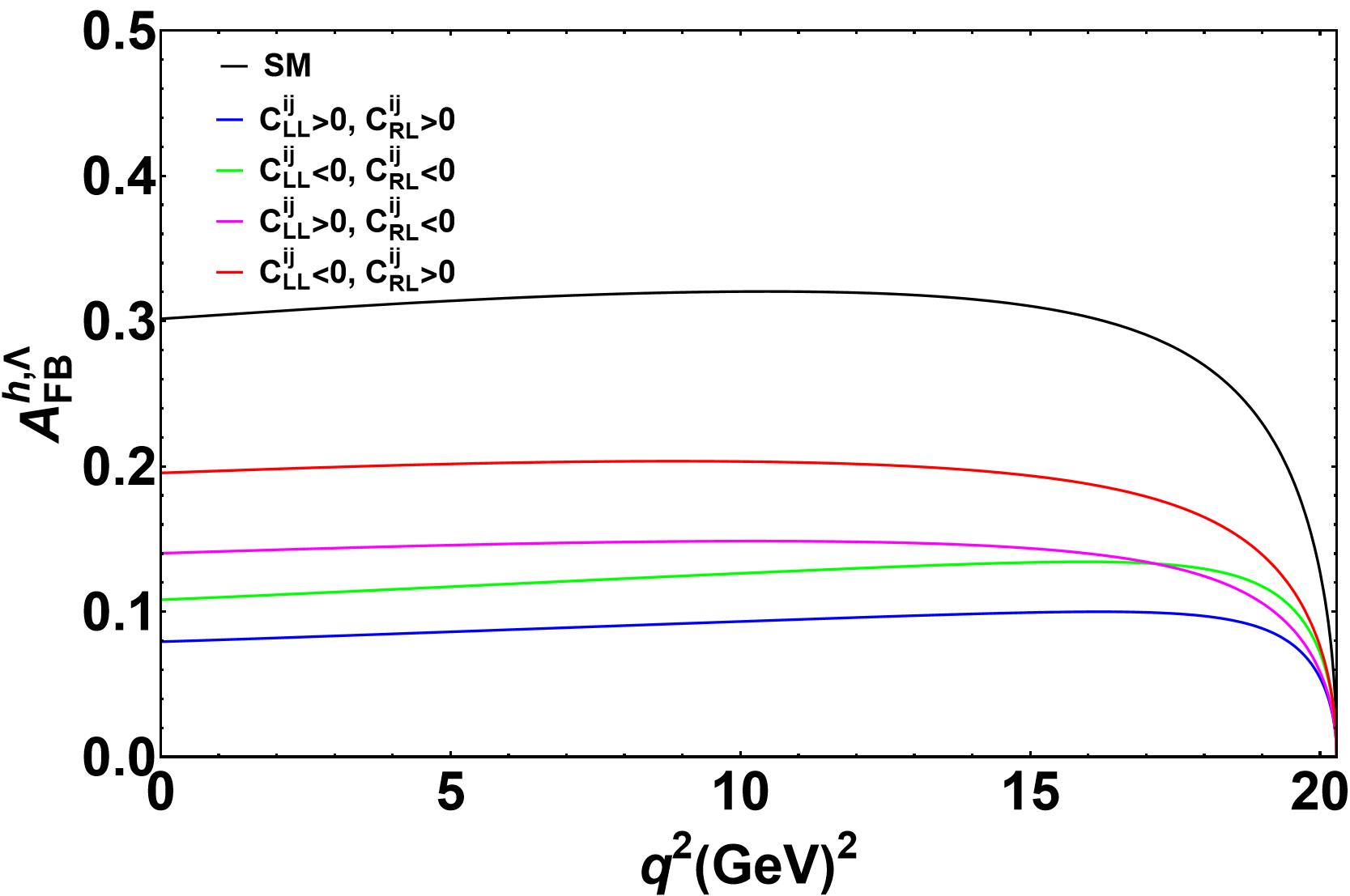}
\caption{The observables $F_L^{\Lambda^\ast}$ and $A_{\rm FB}^{h,\Lambda}$ in the scenario with lepton flavor violation. The first two plots from the left correspond to scenario where $C^{ii}_{LL}=0$, $C^{ii}_{RL}=0$. The last two plots from the left correspond to scenario where $C^{ii}_{LR}=0$ and $C^{ii}_{RR}=0$.}
\label{fig:LFV-NP}
\end{figure}
%

{\it{Lepton Flavor violation:}} If new physics is lepton flavor violating, there is no interference between the NP and the SM.  We consider the following scenario in which both $C^{ij}_{LR}=0$ and $C^{ij}_{RR}=0$ for all flavors and vary $C^{ij}_{RL}$ and $C^{ij}_{LL}$ for all flavors independently subject to the constraints set by the Belle-II data. The upper limit on branching ratios and upper and lower limits on $R^{\Lambda^{(\ast)}}_{\nu\nu}$ are shown in tables \ref{tab:LFV-sc2}}. To obtain the table \ref{tab:LFV-sc2} the $C^{ij}_{LL}$ and $C^{ij}_{RL}$ are varied for three lepton flavors independently. To prepare the table the uncertainties in the theoretical inputs have been included. The other scenario in which $C^{ij}_{RL}=$ and $C^{ij}_{LL}=0$ yields the same table as can be understood from \eqref{eq:LFV-WC}. In figure \ref{fig:LFV-NP} we show the new physics sensitivity of $F_L^{\Lambda^\ast}$ and $A_{\rm FB}^{h,\Lambda}$. The first two plots from left correspond to scenario in which $C^{ij}_{LL}=0$, $C^{ij}_{RL}=0$ and the last two plots correspond to $C^{ij}_{LR}=0$ and $C^{ij}_{RR}=0$. In each plots the set of non-zero coefficients were randomly varied subject to $R_{K^{\ast}}^{\nu\nu}$ constraints. In all the plots, in addition to the SM shown in black, the other colors indicate different NP scenarios. The benchmark points have been chosen for better distinguishability with the SM line. 

In all the three scenarios we find that $A_{\rm FB}^{h,\Lambda}$ is better suited to distinguishing between the SM and the NP scenarios considered. Since the $g_g^{V,A}\to 0$ due to heavy quark symmetry \cite{Das:2020cpv}, we find that $F_L^{\Lambda^\ast} \to 2/3$ irrespective of the NP scenarios considered.

\section{Summary \label{sec:summary}}
The $b\to s\nu\bar{\nu}$ transitions are theoretically cleaner in comparison with $b\to s\ell^+\ell^-$ transitions and serve as sensitive probe of physics beyond the Standard Model. Recently the Belle-II experiment presented their first measurement of $B^+ \to K^+ \nu\bar{\nu}$ decay which deviates from the Standard Model by 2.7$\sigma$. Motivated by these results, in the present paper, we study the baryonic decay modes $\Lambda_b \to \Lambda(\to p\pi)\nu\bar{\nu}$ and $\Lambda_b \to \Lambda^\ast(\to N\!\bar{K})\nu\bar{\nu}$ which are expected to be observed in the proposed Tera-Z factory such as the FCC-ee. The study is performed in a low-energy effective theory framework with additional right-handed neutrinos. We present upper limits on the branching ratios that are consistent with the recent Belle-II result. Additionally, we find that among the two observables studied -- the longitudinal polarization fraction $F_L^{\Lambda^\ast}$ and the hadronic-side forward-backward asymmetry $A_{\rm FB}^{h,\Lambda}$ -- the later can be used to discriminate between the Standard Model and the New Physics scenarios considered.
%

\section*{Acknowledgment}
DD sincerely thanks Hans-Peter Bucheler for his warm hospitality during a visit to the Institute for Theoretical Physics III, University of Stuttgart. DD also acknowledges support under the SERB SRG grant (Sanction Order No. SRG/2023/001318) and the IIIT Hyderabad Seed Fund (No. IIIT/R\&D Office/Seed-Grant/2021-22/013). RS is supported by the National Natural Science Foundation of
China under Grant Nos. 12475094, 12135006 and 12075097, as well as by the Fundamental
Research Funds for the Central Universities under Grant No. CCNU24AI003.,CCNU19TD012.
\appendix
\section{Hadronic Matrix Elements \label{HMA}}
In the helicity basis \cite{Feldmann:2011xf}, the $\Lambda_b\to \Lambda$ hadronic matrix elements for the vector current is
\begin{align}\label{eq:VAhme1}
\langle \Lambda(k,s_k)|\bar{s}\gamma^\mu b |\Lambda_b(p,s_p)\rangle =& \bar{u}(k,s_k)\Bigg[f^V_t(q^2)(m_{\Lambda_b}-m_\Lambda)\frac{q^\mu}{q^2}\nn\\ +& f^V_0(q^2) \frac{m_{\Lambda_b}+m_\Lambda}{s_+} \{p^\mu + k^\mu  - \frac{q^\mu}{q^2}(m^2_{\Lambda_b} - m_\Lambda^2) \} \nn\\ + & f^V_\perp(q^2) \{ \gamma^\mu - \frac{2m_\Lambda}{s_+}p^\mu - \frac{2m_{\Lambda_b}}{s_+}k^\mu \} \Bigg]u(p,s_p)\, ,
\end{align}
and that for the axial-vector current is
\begin{align}\label{eq:VAhme2}
\langle \Lambda(k,s_k)|\bar{s}\gamma^\mu\gamma_5 b |\Lambda(p,s_p)\rangle =& - \bar{u}(k,s_k) \gamma_5 \Bigg[ f_t^A(q^2) (m_{\Lambda_b} + m_\Lambda) \frac{q^\mu}{q^2} \nn\\ +& f_0^A(q^2) \frac{m_{\Lambda_b} - m_\Lambda}{s_-} \{p^\mu + k^\mu - \frac{q^\mu}{q^2} (m^2_{\Lambda_b} - m_\Lambda^2) \} \nn\\ + & f_\perp^A(q^2) \{\gamma^\mu + \frac{2m_\Lambda}{s_-}p^\mu - \frac{2m_{\Lambda_b}}{s_-}k^\mu \}  \Bigg] u(p,s_p)\, .
\end{align}

For the $\Lambda_b\to\Lambda^\ast$ transition the hadronis matrix elements are \cite{Descotes-Genon:2019dbw}
\begin{align}\label{eq:ffVA1}
\langle\Lambda^\ast(k,s_{\Lambda^\ast}) | \bar s \gamma^\mu b|\Lambda_b(p,s_{\Lambda_b})\rangle=
& \bar u_\alpha(k,s_{\Lambda^\ast})\bigg(p^\alpha\biggl[f_t^V(q^2) (m_{\Lambda_b}-m_{\Lambda^\ast})\frac{q^\mu}{q^2}\nn\\
&+ f_0^V(q^2) \frac{m_{\Lambda_b}+m_{\Lambda^\ast}}{s_+}(p^\mu +k^\mu-\frac{q^\mu}{q^2}(m_{\Lambda_b}^2-m_{\Lambda^\ast}^2))\nn\\
&+ f_\perp^V(q^2)(\gamma^\mu-2\frac{m_{\Lambda^\ast}}{s_+}p^\mu -2\frac{m_{\Lambda_b}}{s_+}k^\mu)\biggr]\nn\\
&+ f_g^V(q^2) \left[g^{\alpha\mu}+m_{\Lambda^\ast}\frac{p^\alpha}{s_-} \left(\gamma^\mu - 2 \frac{k^\mu}{m_{\Lambda^\ast}} +2 \frac{m_{\Lambda^\ast} p^\mu +m_{\Lambda_b} k^\mu}{s_+}\right)\right]\bigg)u(p,s_{\Lambda_b})\,,\\
\langle\Lambda^\ast(k,s_{\Lambda^\ast}) | \bar s \gamma^\mu \gamma ^5 b|\Lambda_b(p,s_{\Lambda_b})\rangle=& -\bar u_\alpha(k,s_{\Lambda^\ast})\gamma^5\bigg(p^\alpha\biggl[
f_t^A(q^2) (m_{\Lambda_b}+m_{\Lambda^\ast})\frac{q^\mu}{q^2}\nn\\
&+ f_0^A(q^2) \frac{m_{\Lambda_b}-m_{\Lambda^\ast}}{s_-}(p^\mu +k^\mu-\frac{q^\mu}{q^2}(m_{\Lambda_b}^2-m_{\Lambda^\ast}^2))\nn\\
&+ f_\perp^A(q^2)(\gamma^\mu+2\frac{m_{\Lambda^\ast}}{s_-}p^\mu -2\frac{m_{\Lambda_b}}{s_-}k^\mu)\biggr]\nn\\
&+ f_g^A(q^2) \left[g^{\alpha\mu}-m_{\Lambda^\ast}\frac{p^\alpha}{s_+} \left(\gamma^\mu + 2 \frac{k^\mu}{m_{\Lambda^\ast}} -2 \frac{m_{\Lambda^\ast} p^\mu -m_{\Lambda_b} k^\mu}{s_-}\right)\right]\bigg)u(p,s_{\Lambda_b})\,.
\end{align}

\end{document}